\documentclass[
  aps,
  prl,
  onecolumn,
  superscriptaddress,
  longbibliography,
  reprint
]{revtex4-2}

\usepackage{amsmath,amssymb,bm}
\usepackage{booktabs}
\usepackage{graphicx}
\usepackage{placeins}
\usepackage{hyperref,commath,physics,braket}
\hypersetup{hidelinks}
\usepackage{tikz}
\usepackage[normalem]{ulem}
\DeclareTextCommand{\DJ}{OT1}{{\fontencoding{T1}\selectfont\DJ}}
\usepackage{bibunits,etoolbox}
\usepackage[capitalize]{cleveref}
\usetikzlibrary{arrows.meta,positioning}

\newcommand{\be}{\begin{equation}}
	\newcommand{\ee}{\end{equation}}
\newcommand{\bea}{\begin{eqnarray}}
	\newcommand{\eea}{\end{eqnarray}}
\newcommand{\ba}{\begin{array}}
	\newcommand{\ea}{\end{array}}

\newcommand{\bl}{\begin{flalign}}
	\newcommand{\enl}{\end{flalign}}

\newcommand{\mc}[1]{\mathcal{#1}}

\newcommand{\proj}[1]{\ket{#1}\bra{#1}}

\renewcommand{\bf}[1]{\mathbf{#1}}

\newcommand{\LETTA}{\mathrm{LETTA}}
\newcommand{\MPS}{\mathrm{MPS}}

\begin{document}
\begin{bibunit}[apsrev4-2]
\title{Leg-Tied Tensor Network States: Entanglement Beyond Virtual Bonds}
		\author{Shuoyi Hu}
		\author{Bing Gu}
		\email{gubing@westlake.edu.cn}
		\affiliation{Department of Chemistry and Department of Physics, Westlake University, Hangzhou, Zhejiang 310030, China}
		\date{September 24, 2026}

\begin{abstract}
We introduce a class of tensor-network states in which physical legs are shared among local tensors, termed leg-tied tensor ans\"atze (LETTA). Physical leg ties encode long-range correlations directly, while a virtual matrix product state (MPS) backbone retains short-range multipartite entanglement.
The linear virtual backbone allows us to develop a deterministic density matrix renormalization group-like variational optimization algorithm using exact contractions over the active tie-boundary sets and local minimization.
We demonstrate the advantages of LETTA for the two-dimensional frustrated $J_1$--$J_2$ Heisenberg model and the three-dimensional transverse-field Ising model.
Our results show that LETTA is substantially more accurate than same-bond-dimension MPS calculations and can typically reach the accuracy of much larger MPS calculations using one order of magnitude fewer variational parameters.
LETTA thus opens the door for explicitly correlated tensor-network states  that can encode long-range correlation beyond virtual bonds.
\end{abstract}

\makeatletter
\@author@finish
\global\let\SIauthors\@AAC@list
\global\let\SIaffiliations\@AFF@list
\global\let\SIaffiliationgroups\@AFG@list
\makeatother
\maketitle

Quantum many-body systems with strong correlations represent one of the main challenges in computational physics and chemistry.
Tensor-network states, which encompass many classes of ans\"atze, provide compact descriptions of strongly correlated quantum states for problems ranging from quantum chemistry and lattice models to quantum field theory \cite{xiang2023, ma2022, vidal2008, verstraete2010, tilloy2019}.   Matrix product states (MPS), tensor-network states with a linear chain geometry, are particularly effective in one dimension, where the density matrix renormalization group (DMRG) gives a controlled optimization over the MPS manifold \cite{White1992,Schollwoeck2011,Orus2014, chan2011, schollwock2005}.
In an MPS, correlations are transmitted through virtual bonds, and the bond dimension $D$ controls both expressivity and contraction cost.  The limiting case $D=1$ corresponds to an uncorrelated product state.

The success of MPS and DMRG is deeply related to area laws for entanglement entropy in 1D local, gapped Hamiltonians \cite{eisert2010}. However, applying MPS to two- and higher-dimensional systems or to critical phenomena with long-range correlations remains challenging. For higher-dimensional lattice systems, even short-range Hamiltonians can generate long-range correlation when the system is mapped to a chain. Such long-range correlation cannot be represented efficiently with a finite virtual bond dimension. It is more appropriate to employ the projected entangled-pair states (PEPS), which are natural generalizations of MPS to higher dimensions. However, contracting PEPS is significantly more difficult than MPS, and approximate contraction techniques are being developed \cite{cirac2021, gray2024}.

Correlator product states provide an alternative ansatz for strongly correlated states. For a fixed physical configuration, an MPS amplitude is obtained by contracting auxiliary indices, whereas a correlator product state amplitude is written directly as a product of correlators on overlapping subsets of the physical variables \cite{Changlani2009}. Closely related entangled-plaquette states use plaquette correlators. String-bond states replace correlators with matrix-product amplitudes along selected strings \cite{Schuch2008}. The correlators, rather than a single virtual backbone, determine which physical-variable dependencies are encoded directly. Although such states are commonly optimized by variational Monte Carlo \cite{Mezzacapo2009}, deterministic optimization has been developed for restricted 1D correlator product states \cite{Stojevic2016}.

\begin{figure}[t]
    \centering
    \includegraphics[width=0.48\textwidth]{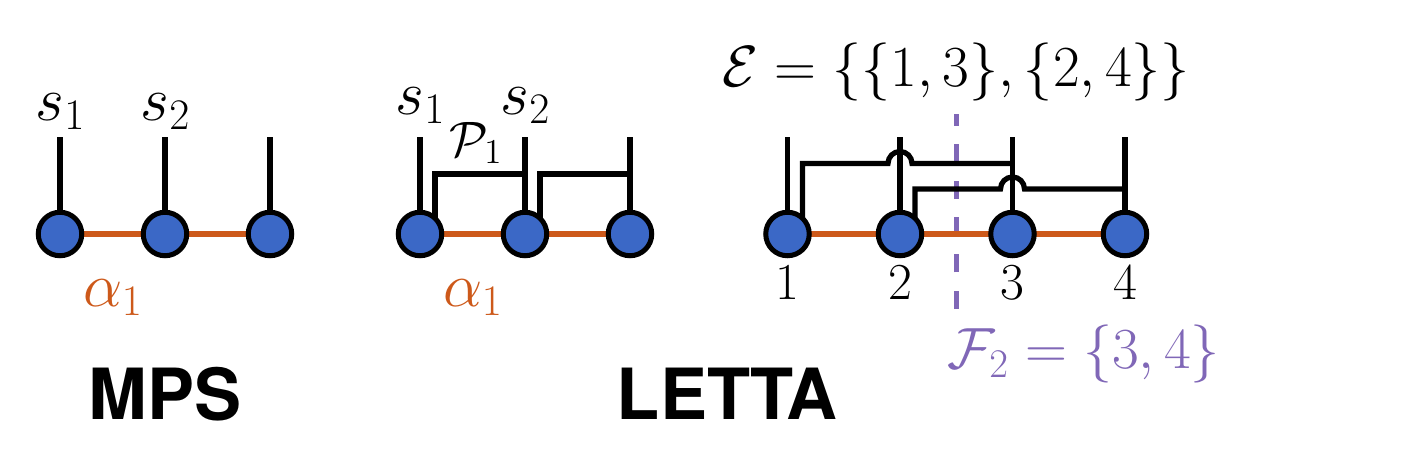}
    \caption{Connectivity patterns of MPS and LETTA. The two LETTA diagrams illustrate nearest-neighbor (NN) and longer-range physical ties. The raised connected bridges mark shared physical legs, and every routed tie is drawn as one continuous path with sharp turns.}
    \label{fig:mps-letta-schematic}
\end{figure}

Here we introduce a class of leg-tied tensor-network states, termed the leg-tied tensor ans\"atze (LETTA). LETTA uses virtual bonds to encode short-range multipartite entanglement like an MPS, but additionally uses explicit ties between physical legs as a direct representation of long-range correlation. Its defining data are therefore an ordered virtual chain and a physical tie graph (Fig.~\ref{fig:mps-letta-schematic}). LETTA contains every MPS with compatible virtual dimensions as a submanifold, whereas at $D=1$ it reduces to a correlator-product or Jastrow amplitude network rather than a product state \cite{Jastrow1955,Changlani2009}. Relative to a correlator product state, LETTA makes the local correlators matrix-valued and contracts their added virtual indices along an MPS backbone. It can thus capture long-range correlation without requiring a large virtual bond dimension needed for an MPS to represent the same structure. This shared-physical-leg structure first appears in the nonadiabatic renormalization group \cite{gu2026}, where neighboring tensors share physical legs generated by the conditional relaxation of the high-energy block with respect to a newly added low-energy site. Unlike a PEPS \cite{Verstraete2008}, LETTA retains one-dimensional ordered virtual bonds, so exact contraction is governed by the tie-boundary width rather than by a two-dimensional virtual network.

We develop a deterministic variational optimization algorithm for LETTA and demonstrate its advantages for the two-dimensional (2D) $J_1$--$J_2$ Heisenberg model and the three-dimensional (3D) transverse-field Ising model. For both models, LETTA is substantially more accurate than the same-bond-dimension MPS, and typically reaches the accuracy of one order-of-magnitude larger bond-dimension MPS calculations.

Consider $N$ ordered sites with local basis states
$s_i\in\{1,\ldots,d_i\}$ and a many-body configuration
$\ket{\mathbf{s}}=\ket{s_1,\ldots,s_N}$.
For an open chain, an MPS is
\begin{equation}
\ket{\Psi_{\MPS}}
= \sum_{\bf s}\sum_{\alpha_1,\ldots,\alpha_{N-1}}
\prod_{i=1}^{N}
B^{[i] s_i}_{\alpha_{i-1}\alpha_i} \ket{\bf s}
\label{eq:mps}
\end{equation}
where $\alpha_i= 0, 1, ..., D_i-1$ are the virtual indices, and $D_i$ is the virtual dimension of the $i$-th bond.
LETTA supplements the virtual chain with a physical tie graph $\mathcal G=(\mc{V},\mc E)$, $\mc{V} =\{1,\ldots,N\}$, as illustrated in Fig.~\ref{fig:mps-letta-schematic}.  Every tie is oriented from the smaller to the larger site label, and $ \mathcal P_i= \left\{ j:\{i,j\}\in \mc{E},\  j > i \right\}$ denotes the future tied neighbors of tensor $i$.
The simplest tie graph for LETTA is the nearest-neighbor tie (middle panel of Fig.~\ref{fig:mps-letta-schematic}) with $\mathcal P_i =\{i+1\}$.
Let $\mathbf{s}_{\mathcal P_i}=(s_j)_{j\in\mathcal P_i}$ denote the corresponding tuple, the LETTA state is then
\begin{equation}
\ket{\Psi}
= \sum_{\bf s} \sum_{\alpha_1,\ldots,\alpha_{N-1}}
\prod_{i=1}^{N}
A^{[i]}_{\alpha_{i-1}\alpha_i}
\bigl(s_{i},\mathbf{s}_{\mathcal P_{i}}\bigr) \ket{\bf s}
\label{eq:letta}
\end{equation}
Thus the physical leg $s_j$ can occur in its own tensor and in every earlier tensor connected to $j$ through edges in $\mc{E}$.  The virtual indices remain ordered as in an MPS, but the physical tie graph can be nonlocal or nonplanar.
The number of stored parameters in LETTA is
\begin{equation}
N_{\LETTA}
=\sum_{i=1}^{N}
D_{i-1}D_i d_{i}
\prod_{j\in\mathcal P_{i}}d_j.
\label{eq:letta-parameter-count}
\end{equation}
For a nearest-neighbor-tied state this is approximately $Nd^2D^2$, versus $NdD^2$ for an MPS. LETTA therefore uses more parameters at the same virtual bond dimension $D$, but places the additional entanglement capacity directly on selected physical correlations.

In the limit of removing all physical ties, i.e., $\mathcal{P}_i=\varnothing~\forall i$, $A^{[i]}_{\alpha_{i-1}\alpha_i} \bigl(s_{i},\mathbf{s}_{\mathcal P_{i}}\bigr) =B^{[i]s_i}_{\alpha_{i-1}\alpha_i}$, LETTA reduces to MPS.
Consequently, at compatible virtual dimensions, the MPS manifold is a submanifold of the LETTA manifold.
In the other limit of removing all virtual bonds (i.e., $D =1$), LETTA reduces to, instead of a product state as for an MPS,  a correlator product state.
For example, without the virtual indices, the nearest-neighbor-tied LETTA then consists of overlapping pair tensors followed by a final one-site tensor, which can be absorbed into its neighbor, giving
$\Psi^{D=1}(\mathbf{s}) = \prod_{i=1}^{N-1}c_i(s_i,s_{i+1}),$
whereas $\Psi_{\MPS}^{D=1}(\mathbf{s})=\prod_{i=1}^{N}b_i(s_i).$
The former is a nearest-neighbor correlator product state; it can carry non-product amplitudes even without virtual bonds. LETTA only becomes a product state when both virtual bonds and physical ties are removed.

\textit{Entanglement entropy $-$}
Dividing the chain into $\mathcal A=\{1,\ldots,m\}$ and its complement $ \bar{\mc{A}}$, the tie-boundary set comprises sites in $\bar{\mathcal A}$ tied to at least one site in $\mathcal A$:
\begin{equation}
\mathcal F_m=
\left(\bigcup_{\ell=1}^{m}\mathcal P_{\ell}\right)
\cap \bar{\mc{A}}
=\left\{j>m:\exists \ell\le m,\ \{\ell,j\}\in\mathcal E\right\}.
\label{eq:cross-cut-tie-set}
\end{equation}
In addition to the virtual bond $\alpha_m$, tensors in $\mathcal A$ can depend on physical legs in $\bar {\mathcal A}$ through $\mathcal F_m$ (see right panel of Fig. 1).
Resolving the configuration $\mathbf q=\mathbf s_{\mathcal F_m}$ gives a left--right factorization of the physical coefficient matrix through the composite boundary $(\alpha_m,\mathbf q)$, of dimension $D_m\prod_{j\in \mathcal F_m}d_j$.  The Schmidt rank obeys
\begin{equation}
r_{\mathcal A}\leq
\min\left\{
\prod_{i\in\mathcal A}d_i,
\prod_{i\in \bar{\mathcal A}}d_i,
D_m\prod_{j\in \mathcal F_m}d_j
\right\}.
\label{eq:letta-schmidt-rank-bound}
\end{equation}
For uniform physical dimension $d$, the corresponding von Neumann entanglement entropy satisfies
\begin{equation}
S_{\mathcal A}\leq
\min\left\{
|\mathcal A|\log d,
|\bar{\mathcal A}|\log d,
\log D_m+|\mathcal F_m|\log d
\right\}.
\label{eq:letta-entanglement-bound}
\end{equation}
The proof is provided in Sec.~\ref{SI-sec:si-entanglement-bound-derivation}.

For an MPS, $\mathcal F_m=\varnothing$, we recover $r_{\mathcal A}\leq D_m$ and $S_{\mathcal A}\leq\log D_m$.  Tied physical states thus provide additional cross-cut channels.

\begin{figure*}[t]
    \centering
    \includegraphics[width=\textwidth]{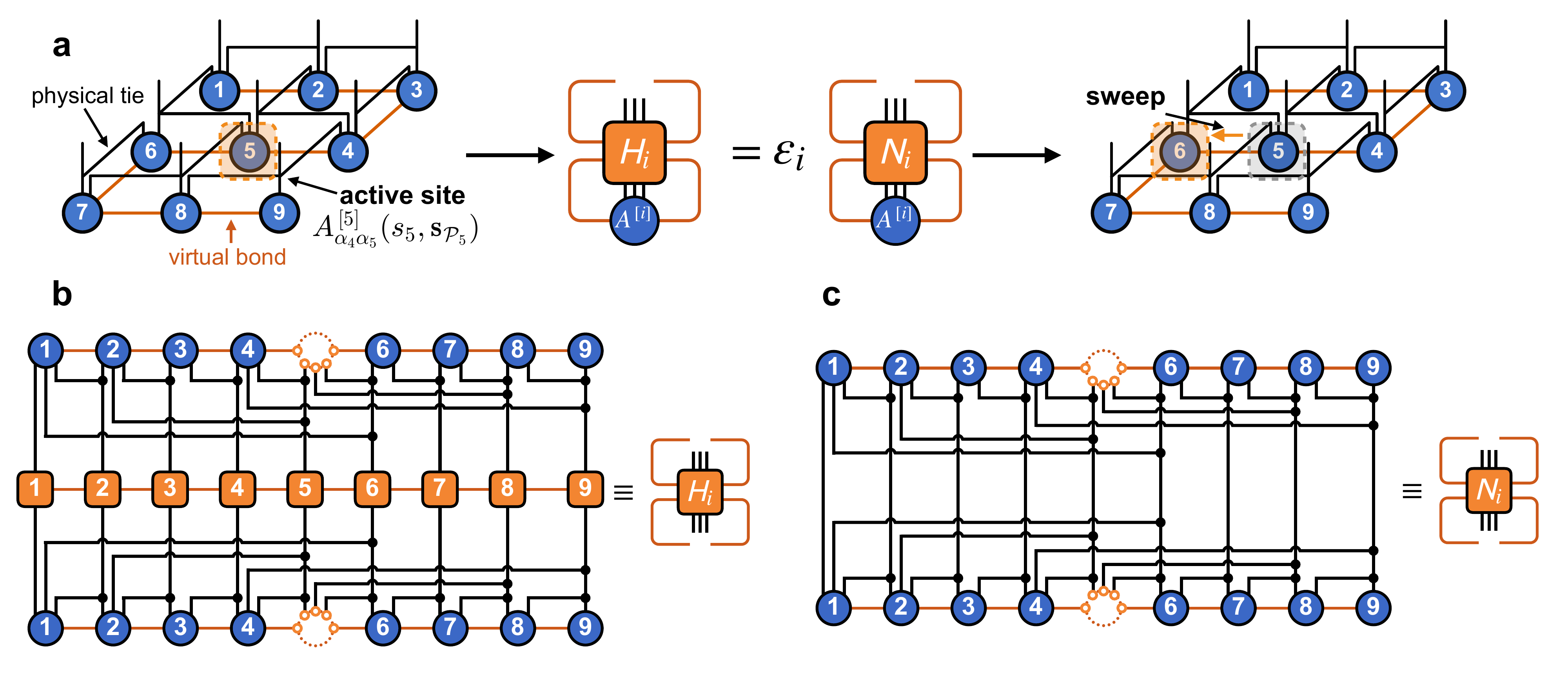}
    \caption{Variational optimization of LETTA.
	(a) At active site $i$, contracting all remaining tensors yields the effective Hamiltonian $\mathsf H_i$ and norm matrix $\mathsf N_i$. The tensor $A^{[i]}$ is updated by solving the generalized eigenvalue problem $\mathsf H_i \mathbf a^{[i]}=\varepsilon_i \mathsf N_i \mathbf a^{[i]}$, after which the sweep advances to site $i+1$.
	(\textbf{b}) and (\textbf{c}) Network contractions defining $\mathsf H_i$ and $\mathsf N_i$. Blue circles denote LETTA tensors, and dotted circles mark the open indices at the active site.}
    \label{fig:frontier-variational-update}
\end{figure*}

\textit{Variational Algorithm ---}
We develop a one-site DMRG-like algorithm to optimize LETTA tensors with the Hamiltonian represented in a matrix product operator (MPO) form. It minimizes the total energy
\begin{equation}
E[\{A\}]=\frac{\langle\Psi|H|\Psi\rangle}{\langle\Psi|\Psi\rangle}
\label{eq:global-rayleigh}
\end{equation}
by alternating one-site updates for a fixed tie graph, ordering, and virtual dimensions.

The central contraction object after sites $1,\ldots,i$ have been processed is the active tie-boundary set $\mathcal F_i$ defined in Eq.~\eqref{eq:cross-cut-tie-set}.
During contraction the number of configurations retained grows as $\prod_{j\in\mathcal F_i}d_j$ for the norm and $\prod_{j\in\mathcal F_i}d_j^2$ for the Hamiltonian expectation.
Thus, the exact contraction avoids the full Hilbert space but remains exponential in the maximum tie-boundary width.

Vectorizing the active tensor as $ \mathbf a^{[i]}=\operatorname{vec}\left(A^{[i]}\right) \in\mathbb C^{n_i}, n_i=D_{i-1}D_i d_i\prod_{j\in\mathcal P_i}d_j$, the wavefunction depends linearly on it
$|\Psi\rangle=\sum_{\mu=1}^{n_i}
a_\mu^{[i]}|\Phi^{[i]}_\mu\rangle,$ where $\mu\leftrightarrow (\alpha_{i-1},s_i,\mathbf s_{\mathcal P_i},\alpha_i)$,
$|\Phi^{[i]}_\mu\rangle$ is obtained by setting $A^{[i]}_\nu=\delta_{\mu\nu}$ and contracting the network with all other tensors fixed.
The local energy is the generalized Rayleigh quotient
\begin{equation}
E_i(\mathbf a^{[i]})
=\frac{\mathbf a^{[i]\dagger}\mathsf H_i\mathbf a^{[i]}}
{\mathbf a^{[i]\dagger}\mathsf N_i\mathbf a^{[i]}}.
\end{equation}
Its stationarity condition yields the generalized eigenvalue problem
\begin{equation}
\mathsf H_i\mathbf a_{\mathrm{new}}^{[i]}
=\varepsilon_i\mathsf N_i\mathbf a_{\mathrm{new}}^{[i]}.
\label{eq:generalized_eigenvalue_eq}
\end{equation}
The local Hamiltonian and norm matrices are $\mathsf H_i=\Phi^{[i]\dagger} H\Phi^{[i]}$ and $\mathsf N_i=\Phi^{[i]\dagger}\Phi^{[i]}$, as shown in Fig.~\ref{fig:frontier-variational-update} b and c, respectively.

\textit{Gauge Fixing ---}
LETTA has gauge freedoms,
\begin{subequations}
\label{eq:LETTA_conditional_canonicalization}
\begin{align}
A^{[m]}(\mathbf q_m)
&\mapsto A^{[m]}(\mathbf q_m)G_m(\mathbf q_m),
\label{eq:LETTA_gauge_left}\\
A^{[m+1]}(\mathbf q_m)
&\mapsto G_m^{-1}(\mathbf q_m)A^{[m+1]}(\mathbf q_m).
\label{eq:LETTA_gauge_right}
\end{align}
\end{subequations}
for each shared physical configuration $\mathbf q_m=\mathbf s_{\mathcal S_m\cap\mathcal S_{m+1}}$, $\mathcal S_i=\{i\}\cup\mathcal P_i$ denotes all physical legs of $A^{[i]}$.
This gauge freedom can be exploited to redistribute scale between neighboring tensors and move an ordinary virtual orthogonality center.
For an MPS, one usually fixes the gauge so that $\mathsf N_m = I$ in Eq.~\eqref{eq:generalized_eigenvalue_eq} \cite{Vidal2003}. This is the so-called mixed-canonical gauge, in which
\begin{equation}
	\sum_{\alpha_{j-1},s_j}
	{B^{[j]s_j*}_{\alpha_{j-1}\alpha_j'}}
	B^{[j]s_j}_{\alpha_{j-1}\alpha_j}
	=\delta_{\alpha_j' \alpha_j},
	\label{eq:mps-left-canonical-conditions}
\end{equation}
for all $j < m$.

For LETTA, we can define a similar conditional canonical gauge with $\mathsf N_m=I$, only when
\begin{equation}
\mathcal F_{m-1}\subseteq\mathcal S_{m-1}\cap\mathcal S_m, \quad
\mathcal F_m\subseteq\mathcal S_m\cap\mathcal S_{m+1}.
\label{eq:conditional_canonical_condition}
\end{equation}
It means physical labels crossing left and right boundaries occur in both adjacent tensors, whereby $\mathbf{q}_m$ in Eq.~\eqref{eq:LETTA_conditional_canonicalization} can be replaced by $\mathbf s_{\mathcal F_m}$. For example, middle panel of Fig.~\ref{fig:mps-letta-schematic} satisfies these conditions while right panel does not. Under these conditions, for each $\mathbf s_{\mathcal F_m}$, we can find a $G_m(\mathbf s_{\mathcal F_m})$ that makes contraction from the right end to site $m$ equal to the identity.

The canonicalization is as follows. Define the right-block states at fixed $\mathbf s_{\mathcal F_m}$ as
\begin{equation}
|r_\alpha(\mathbf s_{\mathcal F_m})\rangle = \sum_{\mathbf s,\boldsymbol\alpha} A^{[m+1]}_{\alpha\alpha_{m+1}} A^{[m+2]}_{\alpha_{m+1}\alpha_{m+2}} \cdots A^{[N]}_{\alpha_{N-1}\alpha_N}
|\mathbf s\rangle,
\end{equation}
where $\mathbf s=\mathbf s_{\bar{\mathcal A}\setminus\mathcal F_m}$, $\boldsymbol\alpha=(\alpha_{m+1},\ldots,\alpha_{N-1})$. With $\mathbf s_{\mathcal F_m}$ held fixed, diagonalize its Gram matrix
$[R_m(\mathbf s_{\mathcal F_m})]_{\alpha\beta} =\langle r_\beta(\mathbf s_{\mathcal F_m})|r_\alpha(\mathbf s_{\mathcal F_m})\rangle$ as $R_m(\mathbf s_{\mathcal F_m})=U\Lambda U^\dagger$, choosing $G_m=U\Lambda^{1/2}$ yields
\begin{equation}
\widetilde R_m(\mathbf s_{\mathcal F_m}) =G_m^{-1}R_m(\mathbf s_{\mathcal F_m})(G_m^{-1})^\dagger=I.
\label{eq:ct-mixed-canonical-conditions}
\end{equation}
The left block can be orthonormalized analogously, yielding $\widetilde L_{m-1}(\mathbf s_{\mathcal F_{m-1}})=I$. With these two conditions, $\widetilde{\mathsf N}_m=I$, such that the contraction and optimization can both be simplified.

When the conditions in Eq.~\eqref{eq:conditional_canonical_condition} are not satisfied, we can employ a tied-leg-independent gauge transformation $G_m\in\mathrm{GL}(D_m)$ to impose

\begin{subequations}
\label{eq:LETTA_virtual_canonicalization}
\begin{align}
\sum_{\alpha_{j-1}, s_j, \mathbf s_{\mathcal P_j}}
\widetilde A^{[j]*}_{\alpha_{j-1}\alpha_j}
\widetilde A^{[j]}_{\alpha_{j-1}\alpha_j'}
&=\delta_{\alpha_j\alpha_j'},
\label{eq:letta-virtual-qr}\\
\sum_{\alpha_j, s_j, \mathbf s_{\mathcal P_j}}
\widetilde A^{[j]}_{\alpha_{j-1}\alpha_j}
\widetilde A^{[j]*}_{\alpha_{j-1}'\alpha_j}
&=\delta_{\alpha_{j-1}\alpha_{j-1}'},
\label{eq:letta-virtual-lq}
\end{align}
\end{subequations}

through QR and LQ factorizations.
These conditions orthogonalize the local tensor unfolding by summing over all physical arguments, which, nevertheless, does not make $\mathsf N_i=I$. Figure~\ref{fig:frontier-variational-update} summarizes the contraction and update; see Secs.~\ref{SI-sec:si-optimization} and~\ref{SI-sec:si-gauges} for whitening, contraction costs, and gauge constructions.

\textit{$Z_2$ Symmetry ---}
In an MPS, abelian symmetry can be enforced by matching physical transformations with representations on the virtual bonds \cite{cirac2021,Singh2011U1}.
The same principle can be applied to LETTA, provided that the tied physical labels are transformed consistently across tensors.

Most relevant here is the Ising $Z_2$ symmetry generated by $P=\prod_{j=1}^N X_j$, where $X, Y, Z$ are Pauli matrices and $j$ runs over the entire chain.
For $g>0$, the nondegenerate ground state is parity even because, for any odd state $|\psi\rangle=\sum_{\bf{s}}\psi_{\bf{s}}|\bf{s}\rangle$, the even state $\sum_{\bf{s}}|\psi_{\bf{s}}||\bf{s}\rangle$ has no higher energy \cite{PhysRevB.68.054405}.
This implies $\Psi(\overline{\mathbf{s}})=\Psi(\mathbf{s})$, where $\overline{\mathbf{s}}=(- s_1,\ldots,- s_N)$ is the spin-flipped configuration, with $s_i=\pm1$.
Assign  charges $q_i(\alpha_i)\in\{0,1\}$ to  virtual legs and define $Q_i=\operatorname{diag}_{\alpha_i}(-1)^{q_i(\alpha_i)} \in \mathbb{R}^{D_i\times D_i}$, this symmetry requires
\begin{equation}
A^{[i]}(\bar s_i,\overline{\mathbf{s}}_{\mathcal P_i})
=Q_{i-1}A^{[i]}(s_i,\mathbf{s}_{\mathcal P_i})Q_i,
\label{eq:letta-z2-equivariance}
\end{equation}
with boundaries $Q_0=Q_N=1$ imposed. Eq.~\eqref{eq:letta-z2-equivariance} suffices to give the desired global parity since each $Q_i$ occurs twice and cancels ($Q_i^2=I$).

To preserve Eq.~\eqref{eq:letta-z2-equivariance} during the left-to-right sweep, we perform QR separately within each parity sector. Reshaping $A^{[i]}$ into a matrix $M_i$ with rows $(\alpha_{i-1},s_i,\mathbf{s}_{\mathcal P_i})$ and columns $\alpha_i$, the local $Z_2$ symmetry condition in Eq.~\eqref{eq:letta-z2-equivariance} becomes
\begin{equation}
(Q_{i-1}\otimes\mathsf F_i) M_i=M_iQ_i,
\label{eq:letta-z2-matrix}
\end{equation}
where $\mathsf F_i=X_i\prod_{j\in\mathcal P_i}X_j$ flips both $s_i$ and $\mathbf{s}_{\mathcal P_i}$.
With a sectorwise QR decomposition $M_i=V_i\mathsf{R}_i$, with $V_i$ unitary and $\mathsf{R}_i$ upper triangular, Eq.~\eqref{eq:letta-z2-matrix} leads to
\begin{equation}
(Q_{i-1}\otimes\mathsf F_i)V_i=V_iQ_i,
\qquad [\mathsf{R}_i,Q_i]=0,
\label{eq:letta-z2-qr}
\end{equation}
see Sec.~\ref{SI-sec:si-z2-qr}. Reshaping $V_i$ gives the parity-preserved $A^{[i]}$. The relation $[\mathsf{R}_i,Q_i]=0$ ensures that after absorbing $\mathsf{R}_i$, $A^{[i+1]}$ preserves Eq.~\eqref{eq:letta-z2-equivariance} too. In the right-to-left sweep, the construction is similar with QR replaced by LQ. This procedure is used for the symmetry-adapted Ising model below.

\begin{figure}[t]
	\centering
	\includegraphics[width=0.45\textwidth]{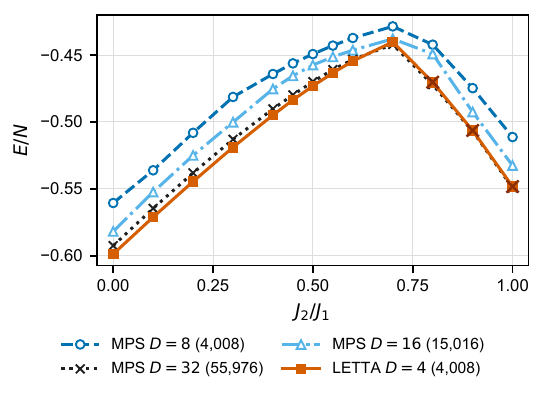}
	\caption{Ground-state energy per site $E/N$ of the $6\times6$ $J_1$--$J_2$ Heisenberg model as a function of $J_2/J_1$ for LETTA $D=4$ and MPS $D=8,16,32$. The numbers in parentheses in the legend indicate the total variational-parameter count for each state. Square-lattice snake site ordering is applied as shown in Fig.~\ref{fig:frontier-variational-update}a.}
	\label{fig:j1j2}
\end{figure}

\textit{2D $J_1$--$J_2$ Heisenberg model ---}
We first consider the spin-$1/2$ $J_1$--$J_2$ Heisenberg model on a square lattice,
\begin{equation}
H_{J_1J_2}
=J_1\sum_{\langle i,j\rangle}\mathbf S_i\cdot\mathbf S_j
+J_2\sum_{\langle\!\langle i,j\rangle\!\rangle}\mathbf S_i\cdot\mathbf S_j,
\label{eq:j1j2-hamiltonian}
\end{equation}
where $\langle i,j\rangle$ and $\langle\!\langle i,j\rangle\!\rangle$ denote nearest- and next-nearest-neighbor pairs, respectively.
This model has been studied extensively with tensor-network methods \cite{YuKao2012,Jiang2012,WangSandvik2018}.

Figure~\ref{fig:frontier-variational-update}a shows the square-lattice site ordering and LETTA connectivity, with virtual bonds following a one-dimensional snake path and physical ties coupling sites across that path. Figure~\ref{fig:j1j2} reports the ground-state energy per site of a $6\times6$ lattice as a function of $J_2/J_1$.  LETTA with $D=4$ is substantially more accurate than MPS $D=4$ (not shown), and yields an even lower energy than the MPS $D=32$ result using only $~7\%$ of its variational parameters. The energy-density maximum near $J_2/J_1\simeq0.7$ is consistent with the reported transition into stripe antiferromagnetic order, estimated at $J_2/J_1\simeq0.62$ in the thermodynamic limit \cite{YuKao2012,Jiang2012}.

\begin{figure*}[t]
    \centering
    \includegraphics[width=0.95\textwidth]{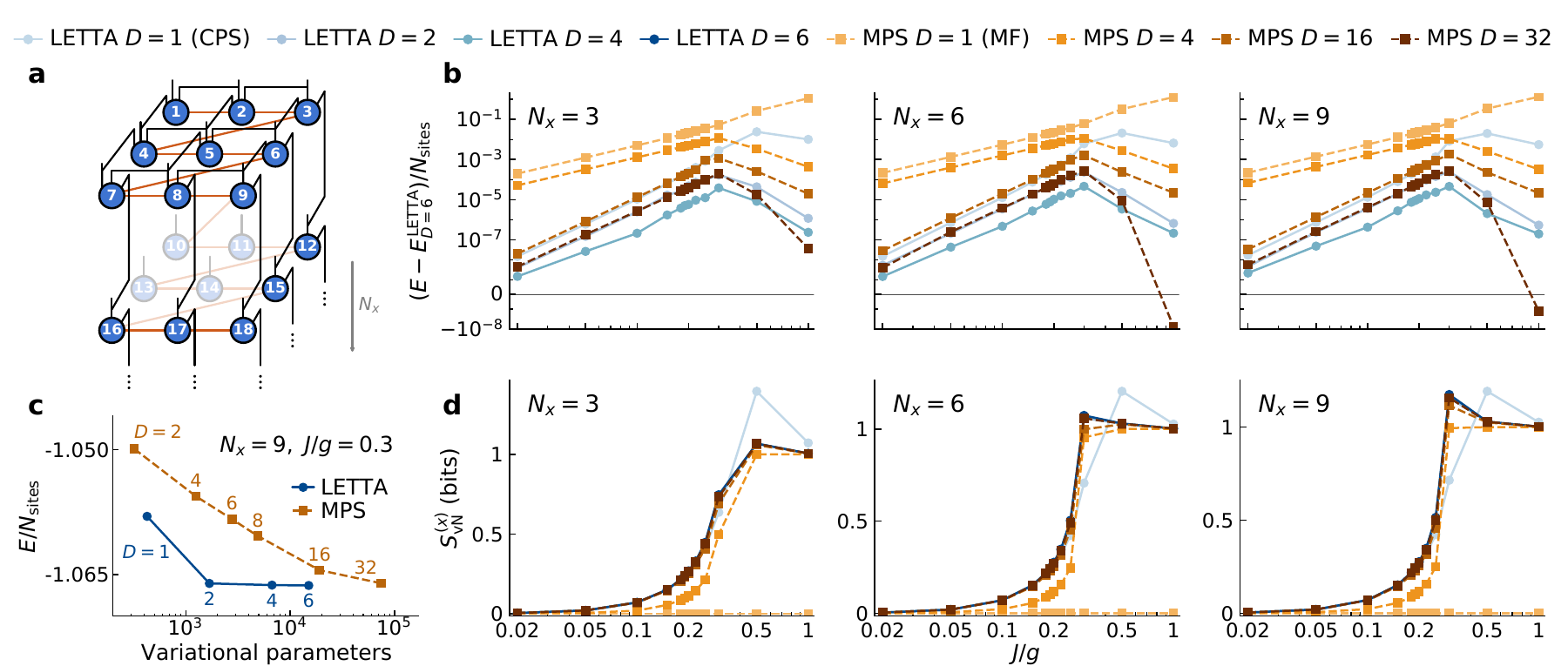}
    \caption{Finite $N_x\times3\times3$ transverse-field Ising
    model studied with LETTA and MPS. (\textbf{a}) Geometry and one-dimensional site ordering used in the LETTA calculations. (\textbf{b}) Energy-density difference
    $(E-E^{\mathrm{LETTA}}_{D=6})/N_{\mathrm{sites}}$.
    (\textbf{c}) Energy density versus the number of variational parameters at $N_x=9$ and $J/g=0.3$, for LETTA $D=1,2,4,6$ and MPS $D=2,4,6,8,16,32$. Counts refer to independent real coefficients in the even-parity sector before removing gauge redundancy.
    (\textbf{d}) Von Neumann entropy $S_{\mathrm{vN}}^{(x)}$ (bits) of the
    normalized stored variational state across the $x$-normal cut after
    $\lfloor N_x/2\rfloor$ layers.  In (b,d), solid blue curves show LETTA $D=1,2,4,6$; dashed orange curves show MPS $D=4,16,32$. CPS: correlator product state, MF: mean-field.}
    \label{fig:3D_ising_energy_difference}
\end{figure*}

\textit{3D transverse-field Ising model ---}
We now consider the 3D $N_x\times3\times3$ clusters of the transverse-field Ising model with Hamiltonian
\begin{equation}
  H_{3D}=-J\sum_{\langle i,j\rangle}Z_i Z_{j}
    -g\sum_iX_i ,
  \label{eq:tfi-hamiltonian}
\end{equation}
where $\langle i,j\rangle$ means all nearest neighbors in the cubic lattice, and we set $g = 1$.
We use one-dimensional raster scanning for the virtual bonds (Fig.~\ref{fig:3D_ising_energy_difference}a) and open boundary conditions in all three directions.

At $J/g<0.3$, we used the gauge in Eq.~\eqref{eq:LETTA_virtual_canonicalization}. These symmetry-unrestricted calculations yield the correct even-parity states ($1-\langle P\rangle<1.1\times10^{-4}$). At $J/g\geq0.3$, we initialize LETTA $D>1$ from an even-parity MPS of the same bond dimension and perform parity-preserving one-site optimization. The $D=1$ LETTA instead starts from a selected parity-preserving correlated state. The MPS results were obtained with two-site DMRG restricted to the even-parity sector.
The convergence criterion  is $\Delta e \leq10^{-7}$, although most cases reach $10^{-8}$.
Details of the local solver, initialization, and continuation are in Sec.~\ref{SI-sec:si-3d-tfim}.

Figure~\ref{fig:3D_ising_energy_difference}b reports the energy-density
difference $(e-e^{\mathrm{LETTA}}_{D=6})$ for $N_x=3,6,9$. Across all computations varying $N_x$ and $J/g$, LETTA yields substantially lower energies than same-bond-dimension MPS.  LETTA $D=4$  yields even lower energy than MPS $D=32$, using only $10\%$ of its  variational parameters.
\cref{fig:3D_ising_energy_difference}c  compares the energy density as a function of the number of variational parameters between LETTA and MPS  at $N_x=9$, $J/g=0.3$.
This clearly shows that LETTA is substantially more efficient as a variational ansatz.
Full parameter counts with and without symmetry are given in Sec.~\ref{SI-sec:si-tfim-parameters} and Tables~\ref{SI-tab:stored-parameter-counts} and~\ref{SI-tab:symmetry-parameter-counts}.

Despite the tied legs, it remains straightforward to compute the entanglement entropy $S = -\Tr \rho_A \log \rho_A $ for LETTA, where $\rho_A = \Tr_{\bar{A}} \proj{\Psi}$ is the reduced density matrix of block $A$. We compute the half-chain entanglement entropy using the Gram-matrix construction described in Sec.~\ref{SI-sec:si-entanglement-spectrum} (see Fig.~\ref{fig:3D_ising_energy_difference}d).
LETTA $S$  essentially converges at the minimal bond dimension $D = 2$, it increases only by $\sim 10^{-3}$  from $D=2$ to $4$.
The LETTA results for $D \ge 2$ all coincide and agree with the MPS $D = 32$ results.
At $J/g\lesssim0.1$, the system is dominated by the field-polarized configuration $\ket{\rightarrow \cdots \rightarrow}$ and can be described easily by MPS as well as LETTA using small bond dimensions.
At $J/g\simeq0.2$--$0.3$, the  entropy rises rapidly.
This is where physical ties provide the most gain in energy, suggesting long-range correlation in the system that is not captured by MPS.  This may also imply a possible quantum critical point, which is consistent with the critical point $(J/g)_c=0.193869(2)$ for the infinite simple-cubic model given by continuous-time cluster Monte Carlo \cite{BloteDeng2002}.
As the coupling strength  reaches $J/g\gtrsim 0.5$, the system is dominated by  two ferromagnetic configurations. The even-parity state $\propto \ket{\uparrow\cdots \uparrow } + \ket{\downarrow \cdots \downarrow}$  has  entropy $S  = 1$ (Fig.~\ref{fig:3D_ising_energy_difference}d).

It is more instructive to examine the entanglement spectra at different coupling regimes ($J/g=0.02$, $0.3$, and $1$), see Fig.~\ref{SI-fig:tfim-entanglement-spectra}.
It reveals that the difficulty here is the broad distribution of entanglement spectra rather than $S$ itself. The long tail at $J/g=0.3$ explains why tied-legs in LETTA provide the most energy gain.
Even at $D= 2$, LETTA already captures many additional Schmidt levels beyond the $D=32$ MPS. These smaller weights collectively contribute to the  entropy, which explains the faster energy and entropy convergence of LETTA at small $D$ compared with MPS.

To summarize, we have introduced a novel type of tensor network states that, in addition to a virtual-bond backbone, employs explicit physical ties to explicitly encode long-range correlations.  The virtual backbone can be contracted and optimized like an MPS. The physical ties allow correlation between far separated sites in the chain, which are crucial for 2D and 3D systems.  We have demonstrated the advantages of LETTA for the 2D $J_1$--$J_2$ Heisenberg and 3D transverse-field Ising models.
To reach the same level of accuracy, LETTA uses substantially fewer variational parameters than MPS. Besides lattice models, we envision many applications of LETTA for problems involving long-range correlation such as quantum chemistry with Coulomb interactions \cite{chan2011, gu_quantum_2025}, vibrational spectroscopy, and disordered systems.
Future perspectives also include developing more robust optimization strategies that can improve the convergence using e.g., quantum Monte Carlo \cite{PhysRevLett.99.220602}.

\begin{acknowledgments}
	This work is supported by the National Natural Science Foundation of China (Grant Nos. 22473090 and 92356310).
\end{acknowledgments}

\nocite{PhysRevA.82.050301}
\putbib[references]
\end{bibunit}

\makeatletter
\def\@extra@b@citeb{SI}
\def\@extra@binfo{SI}
\makeatother
\begin{bibunit}[apsrev4-2]
\onecolumngrid
\clearpage
\setcounter{section}{0}
\setcounter{subsection}{0}
\setcounter{equation}{0}
\setcounter{figure}{0}
\setcounter{table}{0}
\renewcommand{\theHequation}{S\arabic{equation}}
\renewcommand{\theHfigure}{S\arabic{figure}}
\renewcommand{\theHtable}{S\arabic{table}}
\renewcommand{\theHsection}{S\arabic{section}}
\colorlet{RED}{red}
\setcounter{secnumdepth}{3}
\renewcommand{\thesection}{S\arabic{section}}
\renewcommand{\thesubsection}{\thesection\Alph{subsection}}
\renewcommand{\theequation}{S\arabic{equation}}
\renewcommand{\thefigure}{S\arabic{figure}}
\renewcommand{\thetable}{S\arabic{table}}
\makeatletter
\renewcommand{\p@subsection}{}
\setlength{\@fptop}{0pt}
\setlength{\@fpsep}{14pt}
\setlength{\@fpbot}{0pt plus 1fil}
\makeatother

\makeatletter
\ifdefined\titleblock@produce
\begingroup
\let\@AAC@list\SIauthors
\let\@AFF@list\SIaffiliations
\let\@AFG@list\SIaffiliationgroups
\def\@title{Supplemental Material for\texorpdfstring{\\}{ }
Leg-Tied Tensor Network States: Entanglement Beyond Virtual Bonds}
\frontmatter@date{September 24, 2026}
\setbox\absbox\box\voidb@x
\patchcmd{\frontmatter@footnotemark}{frontmatter.#1}{SI-frontmatter.#1}{}{\errmessage{Cannot separate title-note links}}
\patchcmd{\titleblock@produce}{\label{FirstPage}}{\label{SI-FirstPage}}{}{\errmessage{Cannot separate title labels}}
\titleblock@produce
\endgroup
\else
\begin{center}
\textbf{Supplemental Material for\\
Leg-Tied Tensor Network States: Entanglement Beyond Virtual Bonds}\par
Shuoyi Hu and Bing Gu\par
Department of Chemistry and Department of Physics, Westlake University, Hangzhou, Zhejiang 310030, China\par
September 24, 2026\par
\href{mailto:gubing@westlake.edu.cn}{gubing@westlake.edu.cn}
\end{center}
\fi
\makeatother

Section~\ref{SI-sec:si-entanglement-bounds} derives the Schmidt-rank and von Neumann entropy bounds and the entanglement spectrum of the leg-tied tensor ansatz (LETTA); Section~\ref{SI-sec:si-optimization} describes metric-supported optimization and exact contraction of LETTA; Section~\ref{SI-sec:si-gauges} develops LETTA gauge constructions; and Section~\ref{SI-sec:si-3d-tfim} collects the three-dimensional Ising benchmark settings,  entanglement spectra, and parameter counts.

\section{\texorpdfstring{Entanglement bounds and spectrum of LETTA under bipartition}{Entanglement bounds and spectrum of LETTA under a bipartition}}
\label{SI-sec:si-entanglement-bounds}
\subsection{\texorpdfstring{Bipartition}{Bipartition}}
Consider the chain-aligned bipartition $\mathcal A=\{1,\ldots,m\}$ and $\bar{\mathcal A}=\{m+1,\ldots,N_{\mathrm{sites}}\}$ used in Eq.~\eqref{eq:cross-cut-tie-set}. Let $ \mathbf a=(s_i)_{i\in\mathcal A}, \mathbf b=(s_i)_{i\in\bar{\mathcal A}}$ denote complete physical configurations of the two subsystems. Their Hilbert-space dimensions are $d_{\mathcal A}=\prod_{i\in\mathcal A}d_i$ and $d_{\bar{\mathcal A}}=\prod_{i\in\bar{\mathcal A}}d_i$, respectively, with $d_i$ being the physical dimension at site $i$.
The normalized physical coefficient matrix $M\in\mathbb C^{d_{\mathcal A}\times d_{\bar{\mathcal A}}}$ is
\begin{equation}
M_{\mathbf a\mathbf b}
=\frac{\Psi_{\LETTA}(\mathbf a,\mathbf b)}{\sqrt{\mathcal Z}},
\qquad
\mathcal Z=\sum_{\mathbf a,\mathbf b}
|\Psi_{\LETTA}(\mathbf a,\mathbf b)|^2.
\label{SI-eq:normalized-letta-coefficient-matrix}
\end{equation}

The tensors in $\mathcal A$ depend on the physical states in the set of tied sites across the cut $\mathcal F_m$, where $\mathcal F_m=\left(\bigcup_{\ell=1}^{m}\mathcal P_\ell\right)\cap\bar{\mathcal A}$, and $\mathcal P_\ell$ denotes the sites $j>\ell$ tied to site $\ell$.
To separate the contractions on the two sides of the cut, we introduce a copy of the tied physical configuration as a summation index, $\mathbf q=(q_j)_{j\in\mathcal F_m}$, where $q_j\in\{1,\ldots,d_j\}$. With $\alpha_m\in\{0,\ldots,D_m-1\}$ denoting the virtual index crossing the cut, contracting the tensors within each subsystem gives
\begin{equation}
M_{\mathbf a\mathbf b}
=\frac{1}{\sqrt{\mathcal Z}}
\sum_{\alpha_m,\mathbf q}
L_{\alpha_m}(\mathbf a,\mathbf q)\,
\delta_{\mathbf q,\mathbf b_{\mathcal F_m}}\,
R_{\alpha_m}(\mathbf b),
\label{SI-eq:letta-cross-cut-factorization}
\end{equation}
where $\mathbf b_{\mathcal F_m}=(s_j)_{j\in\mathcal F_m}$ is the configuration $\mathbf b$ takes in $\mathcal F_m$.
$\delta_{\mathbf q,\mathbf b_{\mathcal F_m}}$ thus enforces consistency between the boundary configurations taken in both $\mathcal{A}$ and $\bar{\mathcal{A}}$.

Combine the two boundary labels into $\eta=(\alpha_m,\mathbf q)$, whose dimension is $\mathcal D_m=D_m\prod_{j\in\mathcal F_m}d_j.$

Defining
\begin{equation}
\begin{aligned}
\widetilde L_{\mathbf a,\eta}
&=L_{\alpha_m}(\mathbf a,\mathbf q),\\
\widetilde R_{\eta,\mathbf b}
&=\frac{1}{\sqrt{\mathcal Z}}
\delta_{\mathbf q,\mathbf b_{\mathcal F_m}}R_{\alpha_m}(\mathbf b)
\end{aligned}
\label{SI-eq:letta-boundary-factors}
\end{equation}
yields the matrix factorization $M=\widetilde L\widetilde R$, with
$\widetilde L\in\mathbb C^{d_{\mathcal A}\times \mathcal D_m}$ and
$\widetilde R\in\mathbb C^{ \mathcal D_m \times d_{\bar{\mathcal A}}}$.

\subsection{\texorpdfstring{Schmidt-rank and von Neumann entropy bounds}{Schmidt-rank and von Neumann entropy bounds}}
\label{SI-sec:si-entanglement-bound-derivation}
The factorization $M=\widetilde L\widetilde R$ limits the rank of $M$ to the smaller subsystem dimension and the boundary dimension $\mathcal D_m$:
\begin{equation}
\operatorname{rank}M \leq
\min\left\{d_{\mathcal A},d_{\bar{\mathcal A}}, \mathcal D_m \right\} =
\min\left\{
\prod_{i\in\mathcal A}d_i,
\prod_{i\in\bar{\mathcal A}}d_i,
D_m\prod_{j\in\mathcal F_m}d_j
\right\},
\label{SI-eq:appendix-letta-schmidt-rank-bound}
\end{equation}
giving Eq.~\eqref{eq:letta-schmidt-rank-bound}.

Let $r_{\mathcal A}$ denote the Schmidt rank across the bipartition between $\mathcal A$ and $\bar{\mathcal A}$.
The singular values of $M$ are the Schmidt coefficients of the normalized state, so $r_{\mathcal A}=\operatorname{rank}M$.  Equivalently, $\rho_{\mathcal A}=MM^\dagger$ has rank $r_{\mathcal A}$.  We use base-two logarithms for entropies. Its von Neumann entropy is maximized when its nonzero eigenvalues are uniform, giving
\begin{equation}
S_{\mathcal A} \leq \log_2 r_{\mathcal A} \leq
\min\left\{
\sum_{i\in\mathcal A} \log_2 d_i,
\sum_{i\in\bar{\mathcal A}} \log_2 d_i,
\log_2 D_m+\sum_{j\in\mathcal F_m} \log_2 d_j
\right\}.
\label{SI-eq:appendix-letta-entanglement-bound}
\end{equation}
This expression reduces to Eq.~\eqref{eq:letta-entanglement-bound} for uniform local dimension $d$.

\subsection{\texorpdfstring{Entanglement spectrum}{Entanglement spectrum}}
\label{SI-sec:si-entanglement-spectrum}
The same boundary factorization also gives the entanglement spectrum, expressed as the eigenvalues of the reduced density matrix.
Using the factors in Eq.~\eqref{SI-eq:letta-boundary-factors}, define
$|L_\eta\rangle=\sum_{\mathbf a}\widetilde L_{\mathbf a,\eta}|\mathbf a\rangle$ and $|R_\eta\rangle=\sum_{\mathbf b}\widetilde R_{\eta,\mathbf b}|\mathbf b\rangle$.
The normalized state then reads
\begin{equation}
|\Psi\rangle=\sum_{\eta=1}^{\mathcal D_m}|L_\eta\rangle\otimes|R_\eta\rangle.
\label{SI-eq:tfim-entanglement-boundary}
\end{equation}
These boundary states need not be orthonormal. Their Gram matrices are
\begin{equation}
(G_L)_{\eta\eta'}=\langle L_\eta|L_{\eta'}\rangle,
\qquad
(G_R)_{\eta\eta'}=\langle R_\eta|R_{\eta'}\rangle.
\label{SI-eq:tfim-entanglement-grams}
\end{equation}
Equivalently, $G_L=\widetilde L^\dagger\widetilde L$ and $G_R^{\mathsf T}=\widetilde R\widetilde R^\dagger$ so $\rho_{\mathcal A}=MM^\dagger=\widetilde L G_R^{\mathsf T}\widetilde L^\dagger$.
The nonzero eigenvalues of $\mathsf B\mathsf B^\dagger$ and $\mathsf B^\dagger\mathsf B$ coincide for $\mathsf B=\widetilde L(G_R^{\mathsf T})^{1/2}$.
Thus, the reduced-density-matrix eigenvalues can be obtained from a $\mathcal D_m \times \mathcal D_m$ matrix:
\begin{equation}
K=(G_R^{\mathsf T})^{1/2}G_L(G_R^{\mathsf T})^{1/2},
\qquad
p_k=\frac{\lambda_k(K)}{\operatorname{Tr}K}.
\label{SI-eq:tfim-entanglement-spectrum}
\end{equation}
Here, $\lambda_k(K)$ denotes an eigenvalue of $K$, and $\operatorname{Tr}K=1$ for the normalized factorization above. The normalized eigenvalues $\{p_k\}$, equal to the squared Schmidt coefficients, constitute the entanglement spectrum.
The von Neumann entropy in bits is
\begin{equation}
S_{\mathcal A}=-\sum_{p_k>0}p_k\log_2 p_k.
\label{SI-eq:tfim-entanglement-entropy}
\end{equation}

\section{Local optimization and exact contraction}
\label{SI-sec:si-optimization}

\subsection{Metric-supported local minimization}
\label{SI-sec:si-metric}

A local update minimizes the energy while all tensors except $A^{[i]}$ are held fixed.
Write its entries as a vector $\mathbf a^{[i]}\in\mathbb C^{n_i}$, where $n_i$ is the number of entries in the active tensor.
Contracting the remaining tensors gives the effective Hamiltonian $\mathsf H_i$ and norm matrix $\mathsf N_i$, which define the local energy
\begin{equation*}
E_i(\mathbf a^{[i]})=
\frac{\mathbf a^{[i]\dagger}\mathsf H_i\mathbf a^{[i]}}
{\mathbf a^{[i]\dagger}\mathsf N_i\mathbf a^{[i]}}.
\end{equation*}
Tied physical labels and virtual gauge freedom can make the local norm matrix $\mathsf N_i$ singular or poorly conditioned.  To identify directions with vanishing or numerically negligible norm, we diagonalize
\begin{equation}
\mathsf N_i
=U_i\operatorname{diag}(\nu_1,\ldots,\nu_{n_i})U_i^\dagger
\end{equation}
and retain eigenvectors satisfying
\begin{equation}
\mathcal I_i=\{\mu: \nu_\mu>\tau_{\mathrm N} \nu_{\max,i}\},
\qquad
\nu_{\max,i}=\lambda_{\max}(\mathsf N_i),
\qquad
\tau_{\mathrm N}=10^{-12}.
\label{SI-eq:metric-support}
\end{equation}
To make the norm matrix the identity within the retained subspace, we construct the whitener
\begin{equation}
\mathsf X_i=U_{i,\mathcal I_i}
\operatorname{diag}(\nu_\mu^{-1/2})_{\mu\in\mathcal I_i},
\end{equation}
which satisfies $\mathsf X_i^\dagger\mathsf N_i\mathsf X_i=I$. Numerically, we retain the computed projected overlap for stability. The local update is therefore obtained from
\begin{equation}
(\mathsf X_i^\dagger\mathsf H_i\mathsf X_i)\mathbf y_i
=\varepsilon_i (\mathsf X_i^\dagger\mathsf N_i\mathsf X_i)\mathbf y_i,
\qquad
\mathbf a_{\mathrm{new}}^{[i]}=\mathsf X_i \mathbf y_i,
\label{SI-eq:whitened-local-problem}
\end{equation}
using the lowest eigenpair.  The vector $\mathbf y_i$ contains the coordinates in this retained, normalized basis. This restriction removes null parameter directions.

\subsection{Environment updates, acceptance, and contraction cost}
\label{SI-sec:si-contraction}

An environment tensor is a partial contraction of the norm or energy network in Fig.~\ref{fig:frontier-variational-update}b,c of the main text but only on one side of the active site.
During a rightward sweep, environment tensors on the left-hand side are cached, whereas the right-hand-side environment is updated after each local update.
The norm and energy are recomputed by exact contraction over the active tie-boundary set after each full pass.  A proposed tensor is retained only when its local Rayleigh quotient is finite and non-increasing within the numerical tolerance specified for the corresponding benchmark.

The exact contraction remains exponential in the maximum tie-boundary width.
For uniform physical dimension $d$, a Hamiltonian environment at cut $i$ contains $O(D_i^2 \chi_i d^{2|\mathcal F_i|})$ elements. Here, $\chi_i$ is the bond dimension of the Hamiltonian matrix product operator (MPO) at cut $i$.  At fixed tie-boundary width, the number of cached environments grows linearly with the chain length.

\section{Gauge constructions}
\label{SI-sec:si-gauges}

\subsection{Conditional canonicalization of LETTA blocks}
\label{SI-app:conditional-gauge}

For an MPS, the mixed-canonical gauge makes the local norm matrix at the orthogonality center equal to the identity.
For LETTA, the QR/LQ gauge conditions in Eqs.~\eqref{eq:letta-virtual-qr} and~\eqref{eq:letta-virtual-lq} of the main text do not generally yield $\mathsf N_i=I$.
Here, we establish sufficient conditions for LETTA to achieve $\mathsf N_i=I$ by conditionally orthonormalizing the blocks on both sides of the active site.
Although these stronger conditions are not met in the two numerical examples studied here, they can simplify the local eigenproblem and reduce the optimization cost when applicable. We derive the resulting canonical conditions in Eq.~\eqref{eq:ct-mixed-canonical-conditions} of the main text and describe their enforcement during a sweep.
Let $\mathcal S_i=\{i\}\cup\mathcal P_i$ denote the physical labels carried by tensor $A^{[i]}$.
For an interior center at $m$, sufficient boundary conditions are
\begin{equation*}
\mathcal F_{m-1}\subseteq\mathcal S_{m-1}\cap\mathcal S_{m},
\qquad
\mathcal F_{m}\subseteq\mathcal S_{m}\cap\mathcal S_{m+1}.
\end{equation*}
It is equivalent to say that each physical label crossing either boundary occurs in both tensors adjacent to that boundary.
At boundary $m$, let $\mathbf s_{\mathrm L}=(s_1,\ldots,s_m)$ and $\mathbf s_{\mathrm R}=(s_j)_{j>m,\,j\notin\mathcal F_m}$ collect the physical configurations internal to the left and right blocks.
The shared labels $\mathbf s_{\mathcal F_m}$ remain fixed.
At a cut $m$, the block kets $|\ell_\alpha(\mathbf s_{\mathcal F_m})\rangle$ and $|r_\alpha(\mathbf s_{\mathcal F_m})\rangle$ have the coefficients $L_\alpha^{[m]}$ and $R_\alpha^{[m]}$ defined below.
For an active site $m$, the left and right cuts are $m-1$ and $m$, respectively.
Contracting each block gives $L_{\alpha_m}^{[m]}(\mathbf s_{\mathrm L};\mathbf s_{\mathcal F_m})$ and $R_{\alpha_m}^{[m]}(\mathbf s_{\mathrm R};\mathbf s_{\mathcal F_m})$.
We arrange these coefficients into matrices with the virtual index in the columns of $L^{[m]}$ and the rows of $R^{[m]}$:
\begin{equation*}
\begin{aligned}
\bigl[L^{[m]}(\mathbf s_{\mathcal F_m})\bigr]_{\mathbf s_{\mathrm L},\alpha_m}
&=L_{\alpha_m}^{[m]}(\mathbf s_{\mathrm L};\mathbf s_{\mathcal F_m}),\\
\bigl[R^{[m]}(\mathbf s_{\mathcal F_m})\bigr]_{\alpha_m,\mathbf s_{\mathrm R}}
&=R_{\alpha_m}^{[m]}(\mathbf s_{\mathrm R};\mathbf s_{\mathcal F_m}).
\end{aligned}
\end{equation*}
Suppressing the fixed argument $\mathbf s_{\mathcal F_m}$, their conditional Gram matrices are $L_m=\bigl(L^{[m]}\bigr)^\dagger L^{[m]}$ and $R_m=R^{[m]}\bigl(R^{[m]}\bigr)^\dagger$.
As in the main text, $L_m,R_m$ denote the conditional Gram matrices; $L^{[m]},R^{[m]}$ denote the block coefficient matrices.

Since $\mathcal F_m\subseteq\mathcal S_m\cap\mathcal S_{m+1}$, a gauge $G_m(\mathbf s_{\mathcal F_m})$ can act on the two adjacent tensors:
\begin{equation}
\widetilde A^{[m]}=A^{[m]}G_m,\qquad
\widetilde A^{[m+1]}=G_m^{-1}A^{[m+1]}.
\label{SI-eq:conditional-gauge-action}
\end{equation}
The factors cancel at every fixed $\mathbf s_{\mathcal F_m}$, preserving the wavefunction.  For a full-rank left conditional Gram matrix $L_m=U\Lambda U^\dagger>0$, choosing $G_m=U\Lambda^{-1/2}$ gives $\bigl(\widetilde L^{[m]}\bigr)^\dagger\widetilde L^{[m]}=G_m^\dagger L_m G_m=I$.
For a right block with a full-rank conditional Gram matrix $R_m=U\Lambda U^\dagger>0$, the choice $G_m=U\Lambda^{1/2}$ instead gives
$\widetilde R^{[m]}\bigl(\widetilde R^{[m]}\bigr)^\dagger =G_m^{-1} R_m (G_m^{-1})^\dagger=I$.
All matrices depend on the fixed frontier configuration, giving Eq.~\eqref{eq:ct-mixed-canonical-conditions} of the main text.

An invertible gauge preserves rank, so a singular block cannot yield the full identity matrix.  In its eigenbasis, inverse square roots of the positive eigenvalues and finite nonzero scales on null directions give $\operatorname{diag}(I_r,0)$, where $r$ is the block rank.

For nearest-neighbor-tied LETTA, $\mathcal P_j=\{j+1\}$ for $j<N_{\mathrm{sites}}$ and $\mathcal P_{N_{\mathrm{sites}}}=\varnothing$, so $\mathcal F_m=\mathcal S_m\cap\mathcal S_{m+1}=\{m+1\}$ at every internal boundary.  With full-rank conditional block Gram matrices, the canonical conditions around center $m$ take the MPS-like local form
\begin{align}
\sum_{\alpha_{j-1},s_j}
\widetilde A^{[j]*}_{\alpha_{j-1}\alpha_j}
\widetilde A^{[j]}_{\alpha_{j-1}\alpha_j'}
&=\delta_{\alpha_j\alpha_j'},
\label{SI-eq:nn-letta-left-canonical}\\
\sum_{\alpha_j,s_{j+1}}
\widetilde A^{[j]}_{\alpha_{j-1}\alpha_j}
\widetilde A^{[j]*}_{\alpha_{j-1}'\alpha_j}
&=\delta_{\alpha_{j-1}\alpha_{j-1}'},
\label{SI-eq:nn-letta-right-canonical}
\end{align}
The first identity holds for $j<m$ at every fixed $s_{j+1}$, and the second for $m<j<N_{\mathrm{sites}}$ at every fixed $s_j$.  At $j=N_{\mathrm{sites}}$, the same condition uses $\widetilde A^{[N_{\mathrm{sites}}]}(s_{N_{\mathrm{sites}}})$ with no sum over a tied physical label. Conditional QR from the left and LQ from the right establish these identities successively, giving $\widetilde{\mathsf N}_m=I$.
If a conditional block Gram matrix is rank-deficient, the construction above gives the identity on its metric support.

\subsection{Symmetry-preserving QR and LQ}
\label{SI-sec:si-z2-qr}

Decomposing tensors and performing factorizations within symmetry sectors is a standard approach to preserving global symmetries in tensor-network algorithms \cite{PhysRevA.82.050301}.
As an example of symmetry implementation in LETTA, we derive Eq.~\eqref{eq:letta-z2-qr} of the main text by performing QR separately in the two parity sectors.
For a rightward step, reshape $A^{[i]}$ into $M_i$ with rows $(\alpha_{i-1},s_i,\mathbf s_{\mathcal P_i})$ and columns $\alpha_i$.
The operator $Q_i$ assigns parity $+1$ or $-1$ to each virtual basis state; $D_{i,+}$ and $D_{i,-}$ count these states.
The operator $\mathsf F_i$ simultaneously flips all physical spins $(s_i,\mathbf s_{\mathcal P_i})$ carried by tensor $i$.
Group the right virtual indices so that $Q_i=\operatorname{diag}(I_{D_{i,+}},-I_{D_{i,-}})$ and write $M_i=(M_{i,+}\;M_{i,-})$. Equation~\eqref{eq:letta-z2-matrix} of the main text then implies
\begin{equation}
(Q_{i-1}\otimes\mathsf F_i)M_{i,\pm}=\pm M_{i,\pm}.
\end{equation}
Thus the columns of each block lie entirely in the corresponding eigenspace of $Q_{i-1}\otimes\mathsf F_i$.
Let $E_{i,\pm}$ be orthonormal bases of these eigenspaces. Factorizing the coordinates within each eigenspace gives
\begin{equation}
E_{i,\pm}^{\dagger}M_{i,\pm}
=\widehat V_{i,\pm}R_{i,\pm},
\qquad V_{i,\pm}=E_{i,\pm}\widehat V_{i,\pm}.
\end{equation}
Since $(Q_{i-1}\otimes\mathsf F_i)E_{i,\pm}=\pm E_{i,\pm}$, these factors obey $M_{i,\pm}=V_{i,\pm}R_{i,\pm}$ and $(Q_{i-1}\otimes\mathsf F_i)V_{i,\pm}=\pm V_{i,\pm}$.
When a sector has fewer rows than prescribed columns, we pad $V_{i,\pm}$ with zero columns and $R_{i,\pm}$ with zero rows. This preserves the factorization and parity. Assemble
\begin{equation}
V_i=(V_{i,+}\;V_{i,-}),\qquad
R_i=\operatorname{diag}(R_{i,+},R_{i,-}).
\end{equation}
The first identity in Eq.~\eqref{eq:letta-z2-qr} of the main text follows from the definite parity of the columns of $V_i$. The second follows directly from the block structure:
\begin{equation}
R_iQ_i=Q_iR_i=\operatorname{diag}(R_{i,+},-R_{i,-}).
\end{equation}

Finally, let $\mathbf t=(s_{i+1},\mathbf s_{\mathcal P_{i+1}})$ and let $\overline{\mathbf t}$ denote the configuration with every spin flipped, and absorb the transfer factor as $\widetilde A^{[i+1]}(\mathbf t)=R_iA^{[i+1]}(\mathbf t)$.
The commutation relation gives
\begin{align}
\widetilde A^{[i+1]}(\overline{\mathbf t})
&=R_iQ_iA^{[i+1]}(\mathbf t)Q_{i+1}\\
&=Q_i\widetilde A^{[i+1]}(\mathbf t)Q_{i+1}.
\end{align}
Hence both updated tensors retain the symmetry, while $M_iA^{[i+1]}=V_i(R_iA^{[i+1]})$ leaves the contracted state unchanged.
The leftward LQ step follows by the same sectorwise construction.

\section{Three-dimensional transverse-field Ising benchmark}
\label{SI-sec:si-3d-tfim}

\subsection{Geometry, couplings, and tensor dimensions}
\label{SI-sec:si-tfim-geometry}

We use the transverse-field Ising Hamiltonian in main-text Eq.~\eqref{eq:tfi-hamiltonian}, with nearest-neighbor Ising coupling $J$ and transverse field $g$.
$X$, $Y$, and $Z$ denote the single-spin Pauli matrices, with $X$ the spin-flip operator in the $Z$ basis.
We study open $N_x\times3\times3$ clusters with $N_x=3,6,9$, set $g=1$, and sample $J/g=0.02$, $0.05$, $0.10$, $0.15$, $0.18$, $0.19$, $0.20$, $0.22$, $0.25$, $0.30$, $0.50$, and $1.00$.
The sites follow the compact ordering in main-text Fig.~\ref{fig:3D_ising_energy_difference}(a).  Each LETTA tensor contains the current spin, its existing positive-axis neighbors, and two virtual chain legs, giving
\begin{equation}
n_i =D_{i-1}D_i d_i\prod_{j\in\mathcal P_i}d_j, \qquad d_i=2.
\label{SI-eq:tfim-local-dimension}
\end{equation}
For these spin-$1/2$ calculations, we use $Z$-basis labels $s_i=\pm1$, with $Z|s_i\rangle=s_i|s_i\rangle$ and spin flip $\bar s_i=-s_i$. The total number of spins is $N_{\mathrm{sites}}=9N_x$.

\subsection{\texorpdfstring{Optimization, convergence, and initialization}{Optimization, convergence, and initialization}}
\label{SI-sec:si-tfim-convergence}

For symmetry-unrestricted LETTA at $J/g<0.3$, we follow the local optimization procedure in Sec.~\ref{SI-sec:si-optimization} with physical-state-independent virtual QR/LQ. We refine each candidate in the span of the old and proposed tensors, accepting an update only if the local Rayleigh quotient increases by no more than $10^{-9}$.

One full sweep consists of a left-to-right and a right-to-left pass.
The energy-density convergence criterion is $|e^{(p)}-e^{(p-1)}|\leq10^{-7}$, where $e^{(p)}$ is the energy per site after full sweep $p$. Most runs reach $|e^{(p)}-e^{(p-1)}|\leq10^{-8}$. For unrestricted LETTA at $J/g<0.3$, stopping requires a single full sweep with $|e^{(p)}-e^{(p-1)}|\leq10^{-8}$.
For exact-symmetry MPS, the $10^{-7}$ energy-density criterion and a central-cut entropy change below $10^{-6}$ bits hold for three consecutive full sweeps. For symmetry-preserving LETTA, the energy criterion, a parity change below $10^{-7}$, and the absence of rejected local updates hold for three consecutive full sweeps.

For unrestricted LETTA at $J/g<0.3$, at $D=1$, independent random and coupling-continuation starts are optimized, and the lower converged energy is retained.  Higher bond dimensions are generated sequentially along $D=1\rightarrow2\rightarrow4\rightarrow6$ for LETTA.
The lower-$D$ tensors are embedded exactly into the enlarged virtual space, and Gaussian perturbations with $\epsilon_k=5\times10^{-4}2^{-k}$, $k=0,\ldots,5$ (scaled by the root-mean-square entry of each parent tensor) are applied only to newly opened entries, with the lowest-energy candidate selected. The retained $D=1,2,4,6$ states shown in the main text satisfy $1-\langle P\rangle<1.1\times10^{-4}$, with $P=X^{\otimes N_{\mathrm{sites}}}$.

\subsection{MPS reference states}
\label{SI-sec:si-tfim-mps-z2}

All MPS calculations use the even sector of $P$. We label the $X$ eigenbasis by $t\in\{0,1\}$, with $X|t\rangle_x=(-1)^t|t\rangle_x$. Each virtual index carries a binary charge $q_i(\alpha_i)\in\{0,1\}$. The tensors obey
\begin{equation}
B^{[i]t}_{\alpha_{i-1}\alpha_i}=0
\quad\text{unless}\quad
q_{i-1}(\alpha_{i-1})+t=q_i(\alpha_i)\pmod 2,
\end{equation}
with boundary charges $q_0=q_{N_{\mathrm{sites}}}=0$. Two-site density matrix renormalization group (DMRG) optimizes only charge-compatible amplitudes. The optimized tensor is split by a singular-value decomposition in each charge sector; the largest $D$ singular values across both sectors are retained. This allows the sector multiplicities to adapt while preserving total even parity \cite{Schollwoeck2011,cirac2021}.

The main sequence is $D=2\rightarrow4\rightarrow8\rightarrow16\rightarrow32$, starting from a random even-parity MPS. The additional $D=6$ states are initialized from the matching $D=4$ states. Bond expansion retains the existing tensor entries and virtual charges and adds Gaussian noise only to newly opened, charge-compatible entries, with standard deviation $5\times10^{-4}\max(\|B^{[i]}\|_{\mathrm F},1)/\sqrt{2D_{i-1}D_i}$, evaluated using the parent tensor and bond dimensions.

The symmetry-preserving mean-field baseline is evaluated separately at MPS $D=1$. A product state with definite global parity has each spin in an $X$ eigenstate. Its minimum for $g>0$ is $|\rightarrow\rangle^{\otimes N_{\mathrm{sites}}}$, where $X|\rightarrow\rangle=|\rightarrow\rangle$, with $E=-gN_{\mathrm{sites}}$, $\langle P\rangle=1$, and zero cut entropy at every coupling. These analytic results require no optimization sweeps.

\subsection{\texorpdfstring{Initialization and optimization of symmetry-preserving LETTA}{Initialization and optimization of symmetry-preserving LETTA}}
\label{SI-sec:si-tfim-letta-z2}

For the $D=1$ initialization, we use a correlator product state (CPS) \cite{Changlani2009}.

For $J/g=0.3,0.5,1$, the $D=1$ LETTA initial state is $\psi_\kappa(\boldsymbol s)\propto\exp[(\kappa/2)\sum_{\langle i,j\rangle} s_i s_j]$. We choose the lowest-energy candidate from $\kappa\in\{0,0.1,0.2,0.3,0.4,0.6,0.8,1,1.5,2\}$, giving locally spin-flip-invariant $D=1$ tensors, which are subsequently optimized within the full symmetry-preserving tensor space.

For $J/g=0.3,0.5,1$, LETTA with $D=2,4,6$ is initialized from the corresponding even-parity MPS at the same coupling, size, and bond dimension. Transforming the MPS physical index back to the $Z$ basis gives the embedding
\begin{equation}
A^{[i]}_{\alpha_{i-1}\alpha_i}(s_i,\mathbf s_{\mathcal P_i}) =\sum_{t_i=0}^1 W_{s_i t_i}
B^{[i]t_i}_{\alpha_{i-1}\alpha_i}, \qquad W_{st}=\frac{s^t}{\sqrt{2}},\quad s=\pm1,\quad t=0,1.
\end{equation}
Initially the tensor is independent of the tied physical labels $\mathbf s_{\mathcal P_i}$, so this embedding represents the same physical state as the MPS. Its virtual charge labels and occupied sector multiplicities are inherited from the MPS.

Subsequent one-site optimization uses sector-resolved QR/LQ steps, which preserve the symmetry while moving the gauge. The collected states satisfy $|1-\langle P\rangle|<10^{-12}$.

\subsection{\texorpdfstring{Central-cut entanglement spectra}{Central-cut entanglement spectra}}
\label{SI-sec:si-tfim-entropy}
For MPS, the squared normalized Schmidt coefficients give the reduced-density-matrix eigenvalues \cite{Schollwoeck2011}. For LETTA, we use the boundary Gram-matrix construction  in Sec.~\ref{SI-sec:si-entanglement-spectrum}.
For the central $x$-normal cut, the compact ordering divides the system after $\lfloor N_x/2\rfloor$ complete layers, namely after $1$, $3$, and $4$ layers for $N_x=3,6,9$, respectively.
In the site indexing of Sec.~\ref{SI-sec:si-entanglement-bounds}, these cuts correspond to $m=9\lfloor N_x/2\rfloor$. We evaluate $S_{\mathrm{vN}}^{(x)}=S_{\mathcal A}$, denoted by $S$ in the main-text Ising results, using Eqs.~\eqref{SI-eq:tfim-entanglement-spectrum} and~\eqref{SI-eq:tfim-entanglement-entropy}.

\begin{figure}[tbp]
  \centering
  \includegraphics[width=0.75\textwidth]{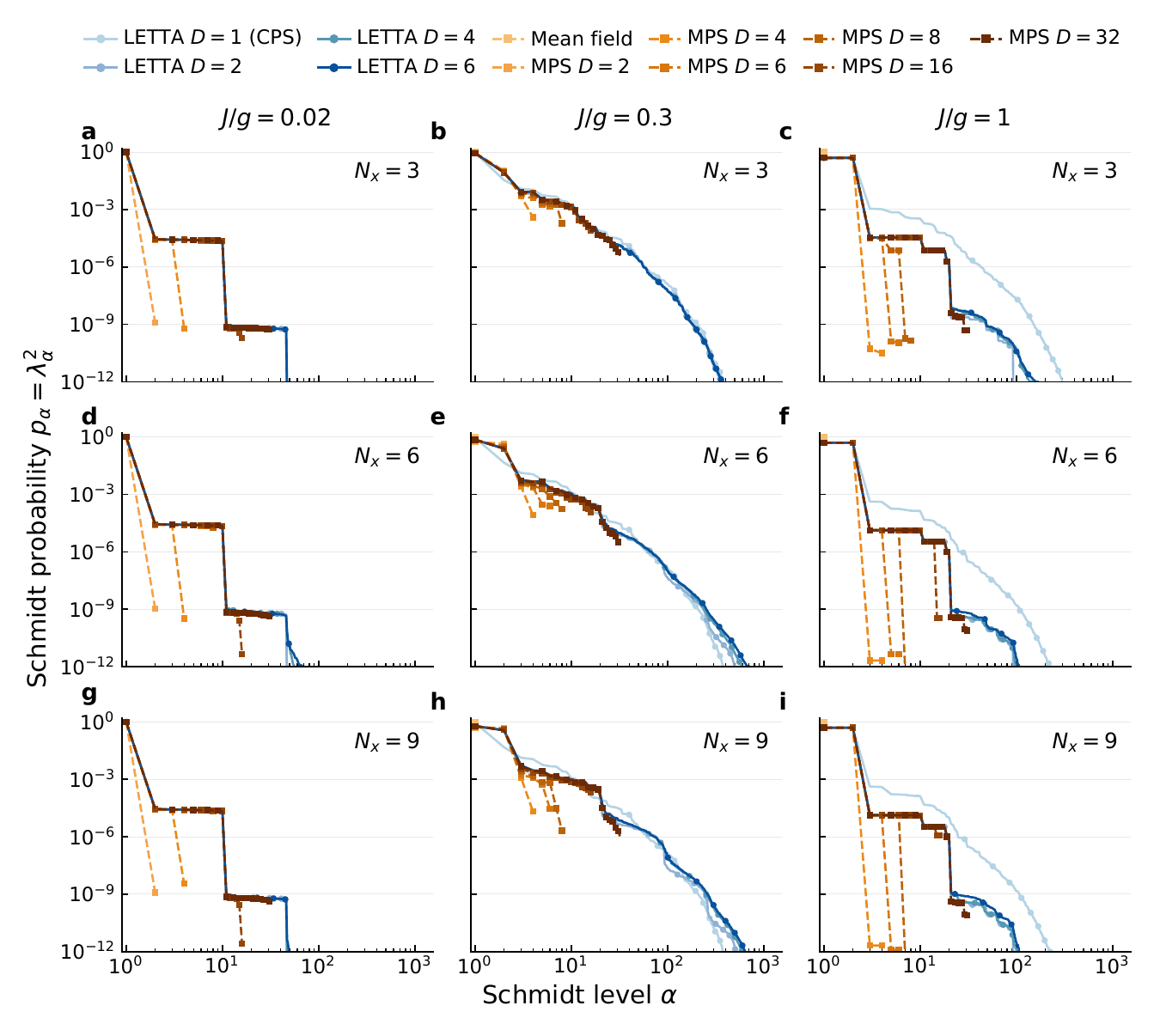}
  \caption{Central-cut entanglement spectra of LETTA and MPS for the
  $N_x\times3\times3$ transverse-field Ising clusters. Rows correspond
  to $N_x=3,6,9$, and columns to $J/g=0.02,0.3,1$.
  The normalized Schmidt probabilities $p_\alpha=\lambda_\alpha^2$ are
  ordered by decreasing weight and plotted on logarithmic axes. Blue
  solid curves show LETTA with $D=1,2,4,6$ ($D=1$ is a correlator product
  state), and orange dashed curves show MPS with $D=2,4,6,8,16,32$ and
  the symmetry-preserving mean-field baseline at $D=1$. The cuts follow
  $1$, $3$, and $4$ complete layers for $N_x=3,6,9$, respectively.
  Only weights above the displayed lower limit of $10^{-12}$ are visible;
  markers indicate a subset of the plotted levels.}
  \label{SI-fig:tfim-entanglement-spectra}
\end{figure}

Figure~\ref{SI-fig:tfim-entanglement-spectra} compares the ordered Schmidt probabilities $p_\alpha=\lambda_\alpha^2$ at three representative couplings.
In the mid-coupling regime, $J/g=0.3$, the LETTA spectra exhibit a broad tail of Schmidt weights extending beyond the rank accessible to the MPS at the bond dimensions considered. An MPS of bond dimension $D$ has at most $D$ nonzero Schmidt weights across this cut, whereas the physical ties allow LETTA to support additional Schmidt levels at the same $D$.

The entropy bound $S_{\mathrm{vN}}^{(x)}\leq\log_2 D$ for an MPS is therefore not a sufficient criterion for representing a target state: even a spectrum with entropy below this bound can contain many nonzero weights beyond its first $D$ levels. For a normalized target spectrum ordered as $p_1\geq p_2\geq\cdots$, truncation to these levels discards the weight
\begin{equation}
\epsilon_D=\sum_{\alpha>D}p_\alpha,
\label{SI-eq:tfim-schmidt-tail}
\end{equation}
Thus a low total entropy does not exclude a loss of information from the many small weights in the tail. The broader spectra at $J/g=0.3$ provide a spectral explanation for the stronger bond-dimension dependence of the MPS results, even when their entropies remain below $\log_2 D$.

\setlength{\textfloatsep}{10pt plus 2pt minus 2pt}
\setlength{\intextsep}{8pt plus 2pt minus 2pt}

\subsection{Variational-parameter counts}
\label{SI-sec:si-tfim-parameters}

Tables~\ref{SI-tab:stored-parameter-counts} and~\ref{SI-tab:symmetry-parameter-counts} count variational parameters, defined as independent real tensor coefficients before removing gauge freedom. The parameter comparison in the main text uses this common convention. With $D_0=D_{N_{\mathrm{sites}}}=1$, the unrestricted counts are $\sum_i2D_{i-1}D_i$ for MPS and $\sum_i2^{1+|\mathcal P_i|}D_{i-1}D_i$ for LETTA. We use $D_i=\min(D,2^i,2^{N_{\mathrm{sites}}-i})$ for MPS and $D_i=D$ on internal LETTA bonds. At fixed tensor shapes, imposing the $\mathbb Z_2$ constraint roughly halves the count for both ans\"atze, leaving their relative parameter reduction unchanged. For MPS, each pair of virtual charges allows one of the two physical states; for LETTA, spin flip pairs the physical configurations and fixes one coefficient from the other.

\begin{table}[htbp]
  \centering
  \caption{Unrestricted variational-parameter counts for the finite
  $N_x\times3\times3$ transverse-field Ising ans\"atze compared in
  main-text Fig.~\ref{fig:3D_ising_energy_difference}.}
  \label{SI-tab:stored-parameter-counts}
  \small
  \setlength{\tabcolsep}{3pt}
  \begin{tabular}{ccrrrrrrr}
    \toprule
    $N_x$ & ansatz & \multicolumn{7}{c}{Maximum bond dimension $D$} \\
    \cmidrule(lr){3-9}
     & & 1 & 2 & 4 & 6 & 8 & 16 & 32 \\
    \midrule
    3 & LETTA & 250 & 964 & 3,784 & 8,460 & -- & -- & -- \\
    3 & MPS & -- & 208 & 776 & 1,648 & 2,856 & 10,408 & 37,544 \\
    \addlinespace
    6 & LETTA & 550 & 2,164 & 8,584 & 19,260 & -- & -- & -- \\
    6 & MPS & -- & 424 & 1,640 & 3,592 & 6,312 & 24,232 & 92,840 \\
    \addlinespace
    9 & LETTA & 850 & 3,364 & 13,384 & 30,060 & -- & -- & -- \\
    9 & MPS & -- & 640 & 2,504 & 5,536 & 9,768 & 38,056 & 148,136 \\
    \bottomrule
  \end{tabular}
\end{table}

\clearpage
\begin{table}[!ht]
  \centering
  \caption{$\mathbb Z_2$-constrained variational-parameter counts for the same ans\"atze and tensor shapes as in Table~\ref{SI-tab:stored-parameter-counts}.}
  \label{SI-tab:symmetry-parameter-counts}
  \small
  \setlength{\tabcolsep}{3pt}

  \begin{tabular}{ccrrrrrrr}
    \toprule
    $N_x$ & ansatz & \multicolumn{7}{c}{Maximum bond dimension $D$} \\
    \cmidrule(lr){3-9}
     & & 1 & 2 & 4 & 6 & 8 & 16 & 32 \\
    \midrule
    3 & LETTA & 125 & 482 & 1,892 & 4,230 & -- & -- & -- \\
    3 & MPS & -- & 104 & 388 & 824 & 1,428 & 5,204 & 18,772 \\
    \addlinespace
    6 & LETTA & 275 & 1,082 & 4,292 & 9,630 & -- & -- & -- \\
    6 & MPS & -- & 212 & 820 & 1,796 & 3,156 & 12,116 & 46,420 \\
    \addlinespace
    9 & LETTA & 425 & 1,682 & 6,692 & 15,030 & -- & -- & -- \\
    9 & MPS & -- & 320 & 1,252 & 2,768 & 4,884 & 19,028 & 74,068 \\
    \bottomrule
  \end{tabular}
\end{table}

\makeatletter
\patchcmd{\present@bibnote}{frontmatter.#1}{SI-frontmatter.#1}{}{\errmessage{Cannot separate author-note anchors}}
\patchcmd{\endthebibliography}{\label{LastBibItem}}{\label{SI-LastBibItem}}{}{\errmessage{Cannot separate bibliography labels}}
\renewcommand{\NAT@bibsetnum}[1]{\setlength{\topsep}{0pt}\NATx@bibsetnum{\ref{SI-LastBibItem}}}
\makeatother
\putbib[references]
\end{bibunit}
\makeatletter\ifdefined\auto@bib@empty\auto@bib@empty\fi\makeatother

\begin{thebibliography}{30}%
\makeatletter
\providecommand \@ifxundefined [1]{%
 \@ifx{#1\undefined}
}%
\providecommand \@ifnum [1]{%
 \ifnum #1\expandafter \@firstoftwo
 \else \expandafter \@secondoftwo
 \fi
}%
\providecommand \@ifx [1]{%
 \ifx #1\expandafter \@firstoftwo
 \else \expandafter \@secondoftwo
 \fi
}%
\providecommand \natexlab [1]{#1}%
\providecommand \enquote  [1]{``#1''}%
\providecommand \bibnamefont  [1]{#1}%
\providecommand \bibfnamefont [1]{#1}%
\providecommand \citenamefont [1]{#1}%
\providecommand \href@noop [0]{\@secondoftwo}%
\providecommand \href [0]{\begingroup \@sanitize@url \@href}%
\providecommand \@href[1]{\@@startlink{#1}\@@href}%
\providecommand \@@href[1]{\endgroup#1\@@endlink}%
\providecommand \@sanitize@url [0]{\catcode `\\12\catcode `\$12\catcode
  `\&12\catcode `\#12\catcode `\^12\catcode `\_12\catcode `\%12\relax}%
\providecommand \@@startlink[1]{}%
\providecommand \@@endlink[0]{}%
\providecommand \url  [0]{\begingroup\@sanitize@url \@url }%
\providecommand \@url [1]{\endgroup\@href {#1}{\urlprefix }}%
\providecommand \urlprefix  [0]{URL }%
\providecommand \Eprint [0]{\href }%
\providecommand \doibase [0]{https://doi.org/}%
\providecommand \selectlanguage [0]{\@gobble}%
\providecommand \bibinfo  [0]{\@secondoftwo}%
\providecommand \bibfield  [0]{\@secondoftwo}%
\providecommand \translation [1]{[#1]}%
\providecommand \BibitemOpen [0]{}%
\providecommand \bibitemStop [0]{}%
\providecommand \bibitemNoStop [0]{.\EOS\space}%
\providecommand \EOS [0]{\spacefactor3000\relax}%
\providecommand \BibitemShut  [1]{\csname bibitem#1\endcsname}%
\let\auto@bib@innerbib\@empty
\bibitem [{\citenamefont {Xiang}(2023)}]{xiang2023}%
  \BibitemOpen
  \bibfield  {author} {\bibinfo {author} {\bibfnamefont {T.}~\bibnamefont
  {Xiang}},\ }\href {https://doi.org/10.1017/9781009398671} {\emph {\bibinfo
  {title} {Density Matrix and Tensor Network Renormalization}}}\ (\bibinfo
  {publisher} {Cambridge University Press},\ \bibinfo {address} {Cambridge},\
  \bibinfo {year} {2023})\BibitemShut {NoStop}%
\bibitem [{\citenamefont {Ma}\ \emph {et~al.}(2022)\citenamefont {Ma},
  \citenamefont {Schollw{\"o}ck},\ and\ \citenamefont {Shuai}}]{ma2022}%
  \BibitemOpen
  \bibfield  {author} {\bibinfo {author} {\bibfnamefont {H.}~\bibnamefont
  {Ma}}, \bibinfo {author} {\bibfnamefont {U.}~\bibnamefont {Schollw{\"o}ck}},\
  and\ \bibinfo {author} {\bibfnamefont {Z.}~\bibnamefont {Shuai}},\ }\href
  {https://doi.org/10.1016/C2020-0-01314-9} {\emph {\bibinfo {title} {Density
  Matrix Renormalization Group ({DMRG})-Based Approaches in Computational
  Chemistry}}},\ \bibinfo {edition} {1st}\ ed.\ (\bibinfo  {publisher}
  {Elsevier},\ \bibinfo {address} {Cambridge, MA},\ \bibinfo {year}
  {2022})\BibitemShut {NoStop}%
\bibitem [{\citenamefont {Vidal}(2008)}]{vidal2008}%
  \BibitemOpen
  \bibfield  {author} {\bibinfo {author} {\bibfnamefont {G.}~\bibnamefont
  {Vidal}},\ }\href {https://doi.org/10.1103/PhysRevLett.101.110501} {\bibfield
   {journal} {\bibinfo  {journal} {Phys. Rev. Lett.}\ }\textbf {\bibinfo
  {volume} {101}},\ \bibinfo {pages} {110501} (\bibinfo {year}
  {2008})}\BibitemShut {NoStop}%
\bibitem [{\citenamefont {Verstraete}\ and\ \citenamefont
  {Cirac}(2010)}]{verstraete2010}%
  \BibitemOpen
  \bibfield  {author} {\bibinfo {author} {\bibfnamefont {F.}~\bibnamefont
  {Verstraete}}\ and\ \bibinfo {author} {\bibfnamefont {J.~I.}\ \bibnamefont
  {Cirac}},\ }\href {https://doi.org/10.1103/PhysRevLett.104.190405} {\bibfield
   {journal} {\bibinfo  {journal} {Phys. Rev. Lett.}\ }\textbf {\bibinfo
  {volume} {104}},\ \bibinfo {pages} {190405} (\bibinfo {year}
  {2010})}\BibitemShut {NoStop}%
\bibitem [{\citenamefont {Tilloy}\ and\ \citenamefont
  {Cirac}(2019)}]{tilloy2019}%
  \BibitemOpen
  \bibfield  {author} {\bibinfo {author} {\bibfnamefont {A.}~\bibnamefont
  {Tilloy}}\ and\ \bibinfo {author} {\bibfnamefont {J.~I.}\ \bibnamefont
  {Cirac}},\ }\href {https://doi.org/10.1103/PhysRevX.9.021040} {\bibfield
  {journal} {\bibinfo  {journal} {Phys. Rev. X}\ }\textbf {\bibinfo {volume}
  {9}},\ \bibinfo {pages} {021040} (\bibinfo {year} {2019})},\ \Eprint
  {https://arxiv.org/abs/1808.00976} {arXiv:1808.00976 [cond-mat.str-el]}
  \BibitemShut {NoStop}%
\bibitem [{\citenamefont {White}(1992)}]{White1992}%
  \BibitemOpen
  \bibfield  {author} {\bibinfo {author} {\bibfnamefont {S.~R.}\ \bibnamefont
  {White}},\ }\href {https://doi.org/10.1103/PhysRevLett.69.2863} {\bibfield
  {journal} {\bibinfo  {journal} {Phys. Rev. Lett.}\ }\textbf {\bibinfo
  {volume} {69}},\ \bibinfo {pages} {2863} (\bibinfo {year}
  {1992})}\BibitemShut {NoStop}%
\bibitem [{\citenamefont {Schollw{\"o}ck}(2011)}]{Schollwoeck2011}%
  \BibitemOpen
  \bibfield  {author} {\bibinfo {author} {\bibfnamefont {U.}~\bibnamefont
  {Schollw{\"o}ck}},\ }\href {https://doi.org/10.1016/j.aop.2010.09.012}
  {\bibfield  {journal} {\bibinfo  {journal} {Ann. Phys. (N.Y.)}\ }\textbf
  {\bibinfo {volume} {326}},\ \bibinfo {pages} {96} (\bibinfo {year}
  {2011})}\BibitemShut {NoStop}%
\bibitem [{\citenamefont {Or{\'u}s}(2014)}]{Orus2014}%
  \BibitemOpen
  \bibfield  {author} {\bibinfo {author} {\bibfnamefont {R.}~\bibnamefont
  {Or{\'u}s}},\ }\href {https://doi.org/10.1016/j.aop.2014.06.013} {\bibfield
  {journal} {\bibinfo  {journal} {Ann. Phys. (N.Y.)}\ }\textbf {\bibinfo
  {volume} {349}},\ \bibinfo {pages} {117} (\bibinfo {year}
  {2014})}\BibitemShut {NoStop}%
\bibitem [{\citenamefont {Chan}\ and\ \citenamefont {Sharma}(2011)}]{chan2011}%
  \BibitemOpen
  \bibfield  {author} {\bibinfo {author} {\bibfnamefont {G.~K.-L.}\
  \bibnamefont {Chan}}\ and\ \bibinfo {author} {\bibfnamefont {S.}~\bibnamefont
  {Sharma}},\ }\href {https://doi.org/10.1146/annurev-physchem-032210-103338}
  {\bibfield  {journal} {\bibinfo  {journal} {Annu. Rev. Phys. Chem.}\ }\textbf
  {\bibinfo {volume} {62}},\ \bibinfo {pages} {465} (\bibinfo {year}
  {2011})}\BibitemShut {NoStop}%
\bibitem [{\citenamefont {Schollw{\"o}ck}(2005)}]{schollwock2005}%
  \BibitemOpen
  \bibfield  {author} {\bibinfo {author} {\bibfnamefont {U.}~\bibnamefont
  {Schollw{\"o}ck}},\ }\href {https://doi.org/10.1103/RevModPhys.77.259}
  {\bibfield  {journal} {\bibinfo  {journal} {Rev. Mod. Phys.}\ }\textbf
  {\bibinfo {volume} {77}},\ \bibinfo {pages} {259} (\bibinfo {year}
  {2005})}\BibitemShut {NoStop}%
\bibitem [{\citenamefont {Eisert}\ \emph {et~al.}(2010)\citenamefont {Eisert},
  \citenamefont {Cramer},\ and\ \citenamefont {Plenio}}]{eisert2010}%
  \BibitemOpen
  \bibfield  {author} {\bibinfo {author} {\bibfnamefont {J.}~\bibnamefont
  {Eisert}}, \bibinfo {author} {\bibfnamefont {M.}~\bibnamefont {Cramer}},\
  and\ \bibinfo {author} {\bibfnamefont {M.~B.}\ \bibnamefont {Plenio}},\
  }\href {https://doi.org/10.1103/RevModPhys.82.277} {\bibfield  {journal}
  {\bibinfo  {journal} {Rev. Mod. Phys.}\ }\textbf {\bibinfo {volume} {82}},\
  \bibinfo {pages} {277} (\bibinfo {year} {2010})}\BibitemShut {NoStop}%
\bibitem [{\citenamefont {Cirac}\ \emph {et~al.}(2021)\citenamefont {Cirac},
  \citenamefont {P{\'e}rez-Garc{\'i}a}, \citenamefont {Schuch},\ and\
  \citenamefont {Verstraete}}]{cirac2021}%
  \BibitemOpen
  \bibfield  {author} {\bibinfo {author} {\bibfnamefont {J.~I.}\ \bibnamefont
  {Cirac}}, \bibinfo {author} {\bibfnamefont {D.}~\bibnamefont
  {P{\'e}rez-Garc{\'i}a}}, \bibinfo {author} {\bibfnamefont {N.}~\bibnamefont
  {Schuch}},\ and\ \bibinfo {author} {\bibfnamefont {F.}~\bibnamefont
  {Verstraete}},\ }\href {https://doi.org/10.1103/RevModPhys.93.045003}
  {\bibfield  {journal} {\bibinfo  {journal} {Rev. Mod. Phys.}\ }\textbf
  {\bibinfo {volume} {93}},\ \bibinfo {pages} {045003} (\bibinfo {year}
  {2021})}\BibitemShut {NoStop}%
\bibitem [{\citenamefont {Gray}\ and\ \citenamefont {Chan}(2024)}]{gray2024}%
  \BibitemOpen
  \bibfield  {author} {\bibinfo {author} {\bibfnamefont {J.}~\bibnamefont
  {Gray}}\ and\ \bibinfo {author} {\bibfnamefont {G.~K.-L.}\ \bibnamefont
  {Chan}},\ }\href {https://doi.org/10.1103/PhysRevX.14.011009} {\bibfield
  {journal} {\bibinfo  {journal} {Phys. Rev. X}\ }\textbf {\bibinfo {volume}
  {14}},\ \bibinfo {pages} {011009} (\bibinfo {year} {2024})}\BibitemShut
  {NoStop}%
\bibitem [{\citenamefont {Changlani}\ \emph {et~al.}(2009)\citenamefont
  {Changlani}, \citenamefont {Kinder}, \citenamefont {Umrigar},\ and\
  \citenamefont {Chan}}]{Changlani2009}%
  \BibitemOpen
  \bibfield  {author} {\bibinfo {author} {\bibfnamefont {H.~J.}\ \bibnamefont
  {Changlani}}, \bibinfo {author} {\bibfnamefont {J.~M.}\ \bibnamefont
  {Kinder}}, \bibinfo {author} {\bibfnamefont {C.~J.}\ \bibnamefont
  {Umrigar}},\ and\ \bibinfo {author} {\bibfnamefont {G.~K.-L.}\ \bibnamefont
  {Chan}},\ }\href {https://doi.org/10.1103/PhysRevB.80.245116} {\bibfield
  {journal} {\bibinfo  {journal} {Phys. Rev. B}\ }\textbf {\bibinfo {volume}
  {80}},\ \bibinfo {pages} {245116} (\bibinfo {year} {2009})}\BibitemShut
  {NoStop}%
\bibitem [{\citenamefont {Schuch}\ \emph {et~al.}(2008)\citenamefont {Schuch},
  \citenamefont {Wolf}, \citenamefont {Verstraete},\ and\ \citenamefont
  {Cirac}}]{Schuch2008}%
  \BibitemOpen
  \bibfield  {author} {\bibinfo {author} {\bibfnamefont {N.}~\bibnamefont
  {Schuch}}, \bibinfo {author} {\bibfnamefont {M.~M.}\ \bibnamefont {Wolf}},
  \bibinfo {author} {\bibfnamefont {F.}~\bibnamefont {Verstraete}},\ and\
  \bibinfo {author} {\bibfnamefont {J.~I.}\ \bibnamefont {Cirac}},\ }\href
  {https://doi.org/10.1103/PhysRevLett.100.040501} {\bibfield  {journal}
  {\bibinfo  {journal} {Phys. Rev. Lett.}\ }\textbf {\bibinfo {volume} {100}},\
  \bibinfo {pages} {040501} (\bibinfo {year} {2008})},\ \Eprint
  {https://arxiv.org/abs/0708.1567} {arXiv:0708.1567 [quant-ph]} \BibitemShut
  {NoStop}%
\bibitem [{\citenamefont {Mezzacapo}\ \emph {et~al.}(2009)\citenamefont
  {Mezzacapo}, \citenamefont {Schuch}, \citenamefont {Boninsegni},\ and\
  \citenamefont {Cirac}}]{Mezzacapo2009}%
  \BibitemOpen
  \bibfield  {author} {\bibinfo {author} {\bibfnamefont {F.}~\bibnamefont
  {Mezzacapo}}, \bibinfo {author} {\bibfnamefont {N.}~\bibnamefont {Schuch}},
  \bibinfo {author} {\bibfnamefont {M.}~\bibnamefont {Boninsegni}},\ and\
  \bibinfo {author} {\bibfnamefont {J.~I.}\ \bibnamefont {Cirac}},\ }\href
  {https://doi.org/10.1088/1367-2630/11/8/083026} {\bibfield  {journal}
  {\bibinfo  {journal} {New J. Phys.}\ }\textbf {\bibinfo {volume} {11}},\
  \bibinfo {pages} {083026} (\bibinfo {year} {2009})},\ \Eprint
  {https://arxiv.org/abs/0905.3898} {arXiv:0905.3898 [cond-mat.str-el]}
  \BibitemShut {NoStop}%
\bibitem [{\citenamefont {Stojevic}\ \emph {et~al.}(2016)\citenamefont
  {Stojevic}, \citenamefont {Crowley}, \citenamefont {{\DJ}uri{\'c}},
  \citenamefont {Grey},\ and\ \citenamefont {Green}}]{Stojevic2016}%
  \BibitemOpen
  \bibfield  {author} {\bibinfo {author} {\bibfnamefont {V.}~\bibnamefont
  {Stojevic}}, \bibinfo {author} {\bibfnamefont {P.}~\bibnamefont {Crowley}},
  \bibinfo {author} {\bibfnamefont {T.}~\bibnamefont {{\DJ}uri{\'c}}}, \bibinfo
  {author} {\bibfnamefont {C.}~\bibnamefont {Grey}},\ and\ \bibinfo {author}
  {\bibfnamefont {A.~G.}\ \bibnamefont {Green}},\ }\href
  {https://doi.org/10.1103/PhysRevB.94.165135} {\bibfield  {journal} {\bibinfo
  {journal} {Phys. Rev. B}\ }\textbf {\bibinfo {volume} {94}},\ \bibinfo
  {pages} {165135} (\bibinfo {year} {2016})},\ \Eprint
  {https://arxiv.org/abs/1604.07210} {arXiv:1604.07210 [quant-ph]} \BibitemShut
  {NoStop}%
\bibitem [{\citenamefont {Jastrow}(1955)}]{Jastrow1955}%
  \BibitemOpen
  \bibfield  {author} {\bibinfo {author} {\bibfnamefont {R.}~\bibnamefont
  {Jastrow}},\ }\href {https://doi.org/10.1103/PhysRev.98.1479} {\bibfield
  {journal} {\bibinfo  {journal} {Phys. Rev.}\ }\textbf {\bibinfo {volume}
  {98}},\ \bibinfo {pages} {1479} (\bibinfo {year} {1955})}\BibitemShut
  {NoStop}%
\bibitem [{\citenamefont {Gu}(2026)}]{gu2026}%
  \BibitemOpen
  \bibfield  {author} {\bibinfo {author} {\bibfnamefont {B.}~\bibnamefont
  {Gu}},\ }\href {https://doi.org/10.48550/arXiv.2604.27381} {\bibinfo {title}
  {Nonadiabatic renormalization group for strongly coupled multiscale quantum
  systems}} (\bibinfo {year} {2026}),\ \Eprint
  {https://arxiv.org/abs/2604.27381} {arXiv:2604.27381 [quant-ph]} \BibitemShut
  {NoStop}%
\bibitem [{\citenamefont {Verstraete}\ \emph {et~al.}(2008)\citenamefont
  {Verstraete}, \citenamefont {Murg},\ and\ \citenamefont
  {Cirac}}]{Verstraete2008}%
  \BibitemOpen
  \bibfield  {author} {\bibinfo {author} {\bibfnamefont {F.}~\bibnamefont
  {Verstraete}}, \bibinfo {author} {\bibfnamefont {V.}~\bibnamefont {Murg}},\
  and\ \bibinfo {author} {\bibfnamefont {J.~I.}\ \bibnamefont {Cirac}},\ }\href
  {https://doi.org/10.1080/14789940801912366} {\bibfield  {journal} {\bibinfo
  {journal} {Adv. Phys.}\ }\textbf {\bibinfo {volume} {57}},\ \bibinfo {pages}
  {143} (\bibinfo {year} {2008})}\BibitemShut {NoStop}%
\bibitem [{\citenamefont {Vidal}(2003)}]{Vidal2003}%
  \BibitemOpen
  \bibfield  {author} {\bibinfo {author} {\bibfnamefont {G.}~\bibnamefont
  {Vidal}},\ }\href {https://doi.org/10.1103/PhysRevLett.91.147902} {\bibfield
  {journal} {\bibinfo  {journal} {Phys. Rev. Lett.}\ }\textbf {\bibinfo
  {volume} {91}},\ \bibinfo {pages} {147902} (\bibinfo {year}
  {2003})}\BibitemShut {NoStop}%
\bibitem [{\citenamefont {Singh}\ \emph {et~al.}(2011)\citenamefont {Singh},
  \citenamefont {Pfeifer},\ and\ \citenamefont {Vidal}}]{Singh2011U1}%
  \BibitemOpen
  \bibfield  {author} {\bibinfo {author} {\bibfnamefont {S.}~\bibnamefont
  {Singh}}, \bibinfo {author} {\bibfnamefont {R.~N.~C.}\ \bibnamefont
  {Pfeifer}},\ and\ \bibinfo {author} {\bibfnamefont {G.}~\bibnamefont
  {Vidal}},\ }\href {https://doi.org/10.1103/PhysRevB.83.115125} {\bibfield
  {journal} {\bibinfo  {journal} {Phys. Rev. B}\ }\textbf {\bibinfo {volume}
  {83}},\ \bibinfo {pages} {115125} (\bibinfo {year} {2011})}\BibitemShut
  {NoStop}%
\bibitem [{\citenamefont {Moessner}\ and\ \citenamefont
  {Sondhi}(2003)}]{PhysRevB.68.054405}%
  \BibitemOpen
  \bibfield  {author} {\bibinfo {author} {\bibfnamefont {R.}~\bibnamefont
  {Moessner}}\ and\ \bibinfo {author} {\bibfnamefont {S.~L.}\ \bibnamefont
  {Sondhi}},\ }\href {https://doi.org/10.1103/PhysRevB.68.054405} {\bibfield
  {journal} {\bibinfo  {journal} {Phys. Rev. B}\ }\textbf {\bibinfo {volume}
  {68}},\ \bibinfo {pages} {054405} (\bibinfo {year} {2003})}\BibitemShut
  {NoStop}%
\bibitem [{\citenamefont {Yu}\ and\ \citenamefont {Kao}(2012)}]{YuKao2012}%
  \BibitemOpen
  \bibfield  {author} {\bibinfo {author} {\bibfnamefont {J.-F.}\ \bibnamefont
  {Yu}}\ and\ \bibinfo {author} {\bibfnamefont {Y.-J.}\ \bibnamefont {Kao}},\
  }\href {https://doi.org/10.1103/PhysRevB.85.094407} {\bibfield  {journal}
  {\bibinfo  {journal} {Phys. Rev. B}\ }\textbf {\bibinfo {volume} {85}},\
  \bibinfo {pages} {094407} (\bibinfo {year} {2012})}\BibitemShut {NoStop}%
\bibitem [{\citenamefont {Jiang}\ \emph {et~al.}(2012)\citenamefont {Jiang},
  \citenamefont {Yao},\ and\ \citenamefont {Balents}}]{Jiang2012}%
  \BibitemOpen
  \bibfield  {author} {\bibinfo {author} {\bibfnamefont {H.-C.}\ \bibnamefont
  {Jiang}}, \bibinfo {author} {\bibfnamefont {H.}~\bibnamefont {Yao}},\ and\
  \bibinfo {author} {\bibfnamefont {L.}~\bibnamefont {Balents}},\ }\href
  {https://doi.org/10.1103/PhysRevB.86.024424} {\bibfield  {journal} {\bibinfo
  {journal} {Phys. Rev. B}\ }\textbf {\bibinfo {volume} {86}},\ \bibinfo
  {pages} {024424} (\bibinfo {year} {2012})}\BibitemShut {NoStop}%
\bibitem [{\citenamefont {Wang}\ and\ \citenamefont
  {Sandvik}(2018)}]{WangSandvik2018}%
  \BibitemOpen
  \bibfield  {author} {\bibinfo {author} {\bibfnamefont {L.}~\bibnamefont
  {Wang}}\ and\ \bibinfo {author} {\bibfnamefont {A.~W.}\ \bibnamefont
  {Sandvik}},\ }\href {https://doi.org/10.1103/PhysRevLett.121.107202}
  {\bibfield  {journal} {\bibinfo  {journal} {Phys. Rev. Lett.}\ }\textbf
  {\bibinfo {volume} {121}},\ \bibinfo {pages} {107202} (\bibinfo {year}
  {2018})}\BibitemShut {NoStop}%
\bibitem [{\citenamefont {Bl{\"o}te}\ and\ \citenamefont
  {Deng}(2002)}]{BloteDeng2002}%
  \BibitemOpen
  \bibfield  {author} {\bibinfo {author} {\bibfnamefont {H.~W.~J.}\
  \bibnamefont {Bl{\"o}te}}\ and\ \bibinfo {author} {\bibfnamefont
  {Y.}~\bibnamefont {Deng}},\ }\href
  {https://doi.org/10.1103/PhysRevE.66.066110} {\bibfield  {journal} {\bibinfo
  {journal} {Phys. Rev. E}\ }\textbf {\bibinfo {volume} {66}},\ \bibinfo
  {pages} {066110} (\bibinfo {year} {2002})}\BibitemShut {NoStop}%
\bibitem [{\citenamefont {Gu}\ \emph {et~al.}(2025)\citenamefont {Gu},
  \citenamefont {Ren},\ and\ \citenamefont {Zhang}}]{gu_quantum_2025}%
  \BibitemOpen
  \bibfield  {author} {\bibinfo {author} {\bibfnamefont {B.}~\bibnamefont
  {Gu}}, \bibinfo {author} {\bibfnamefont {J.}~\bibnamefont {Ren}},\ and\
  \bibinfo {author} {\bibfnamefont {J.}~\bibnamefont {Zhang}},\ }\href
  {https://doi.org/10.1021/acs.jctc.5c00029} {\bibfield  {journal} {\bibinfo
  {journal} {J. Chem. Theory Comput.}\ }\textbf {\bibinfo {volume} {21}},\
  \bibinfo {pages} {6793} (\bibinfo {year} {2025})}\BibitemShut {NoStop}%
\bibitem [{\citenamefont {Sandvik}\ and\ \citenamefont
  {Vidal}(2007)}]{PhysRevLett.99.220602}%
  \BibitemOpen
  \bibfield  {author} {\bibinfo {author} {\bibfnamefont {A.~W.}\ \bibnamefont
  {Sandvik}}\ and\ \bibinfo {author} {\bibfnamefont {G.}~\bibnamefont
  {Vidal}},\ }\href {https://doi.org/10.1103/PhysRevLett.99.220602} {\bibfield
  {journal} {\bibinfo  {journal} {Phys. Rev. Lett.}\ }\textbf {\bibinfo
  {volume} {99}},\ \bibinfo {pages} {220602} (\bibinfo {year}
  {2007})}\BibitemShut {NoStop}%
\bibitem [{\citenamefont {Singh}\ \emph {et~al.}(2010)\citenamefont {Singh},
  \citenamefont {Pfeifer},\ and\ \citenamefont {Vidal}}]{PhysRevA.82.050301}%
  \BibitemOpen
  \bibfield  {author} {\bibinfo {author} {\bibfnamefont {S.}~\bibnamefont
  {Singh}}, \bibinfo {author} {\bibfnamefont {R.~N.~C.}\ \bibnamefont
  {Pfeifer}},\ and\ \bibinfo {author} {\bibfnamefont {G.}~\bibnamefont
  {Vidal}},\ }\href {https://doi.org/10.1103/PhysRevA.82.050301} {\bibfield
  {journal} {\bibinfo  {journal} {Phys. Rev. A}\ }\textbf {\bibinfo {volume}
  {82}},\ \bibinfo {pages} {050301(R)} (\bibinfo {year} {2010})}\BibitemShut
  {NoStop}%
\end{thebibliography}%


\begin{thebibliography}{4}%
\makeatletter
\providecommand \@ifxundefined [1]{%
 \@ifx{#1\undefined}
}%
\providecommand \@ifnum [1]{%
 \ifnum #1\expandafter \@firstoftwo
 \else \expandafter \@secondoftwo
 \fi
}%
\providecommand \@ifx [1]{%
 \ifx #1\expandafter \@firstoftwo
 \else \expandafter \@secondoftwo
 \fi
}%
\providecommand \natexlab [1]{#1}%
\providecommand \enquote  [1]{``#1''}%
\providecommand \bibnamefont  [1]{#1}%
\providecommand \bibfnamefont [1]{#1}%
\providecommand \citenamefont [1]{#1}%
\providecommand \href@noop [0]{\@secondoftwo}%
\providecommand \href [0]{\begingroup \@sanitize@url \@href}%
\providecommand \@href[1]{\@@startlink{#1}\@@href}%
\providecommand \@@href[1]{\endgroup#1\@@endlink}%
\providecommand \@sanitize@url [0]{\catcode `\\12\catcode `\$12\catcode
  `\&12\catcode `\#12\catcode `\^12\catcode `\_12\catcode `\%12\relax}%
\providecommand \@@startlink[1]{}%
\providecommand \@@endlink[0]{}%
\providecommand \url  [0]{\begingroup\@sanitize@url \@url }%
\providecommand \@url [1]{\endgroup\@href {#1}{\urlprefix }}%
\providecommand \urlprefix  [0]{URL }%
\providecommand \Eprint [0]{\href }%
\providecommand \doibase [0]{https://doi.org/}%
\providecommand \selectlanguage [0]{\@gobble}%
\providecommand \bibinfo  [0]{\@secondoftwo}%
\providecommand \bibfield  [0]{\@secondoftwo}%
\providecommand \translation [1]{[#1]}%
\providecommand \BibitemOpen [0]{}%
\providecommand \bibitemStop [0]{}%
\providecommand \bibitemNoStop [0]{.\EOS\space}%
\providecommand \EOS [0]{\spacefactor3000\relax}%
\providecommand \BibitemShut  [1]{\csname bibitem#1\endcsname}%
\let\auto@bib@innerbib\@empty
\bibitem [{\citenamefont {Singh}\ \emph {et~al.}(2010)\citenamefont {Singh},
  \citenamefont {Pfeifer},\ and\ \citenamefont {Vidal}}]{PhysRevA.82.050301}%
  \BibitemOpen
  \bibfield  {author} {\bibinfo {author} {\bibfnamefont {S.}~\bibnamefont
  {Singh}}, \bibinfo {author} {\bibfnamefont {R.~N.~C.}\ \bibnamefont
  {Pfeifer}},\ and\ \bibinfo {author} {\bibfnamefont {G.}~\bibnamefont
  {Vidal}},\ }\href {https://doi.org/10.1103/PhysRevA.82.050301} {\bibfield
  {journal} {\bibinfo  {journal} {Phys. Rev. A}\ }\textbf {\bibinfo {volume}
  {82}},\ \bibinfo {pages} {050301(R)} (\bibinfo {year} {2010})}\BibitemShut
  {NoStop}%
\bibitem [{\citenamefont {Schollw{\"o}ck}(2011)}]{Schollwoeck2011}%
  \BibitemOpen
  \bibfield  {author} {\bibinfo {author} {\bibfnamefont {U.}~\bibnamefont
  {Schollw{\"o}ck}},\ }\href {https://doi.org/10.1016/j.aop.2010.09.012}
  {\bibfield  {journal} {\bibinfo  {journal} {Ann. Phys. (N.Y.)}\ }\textbf
  {\bibinfo {volume} {326}},\ \bibinfo {pages} {96} (\bibinfo {year}
  {2011})}\BibitemShut {NoStop}%
\bibitem [{\citenamefont {Cirac}\ \emph {et~al.}(2021)\citenamefont {Cirac},
  \citenamefont {P{\'e}rez-Garc{\'i}a}, \citenamefont {Schuch},\ and\
  \citenamefont {Verstraete}}]{cirac2021}%
  \BibitemOpen
  \bibfield  {author} {\bibinfo {author} {\bibfnamefont {J.~I.}\ \bibnamefont
  {Cirac}}, \bibinfo {author} {\bibfnamefont {D.}~\bibnamefont
  {P{\'e}rez-Garc{\'i}a}}, \bibinfo {author} {\bibfnamefont {N.}~\bibnamefont
  {Schuch}},\ and\ \bibinfo {author} {\bibfnamefont {F.}~\bibnamefont
  {Verstraete}},\ }\href {https://doi.org/10.1103/RevModPhys.93.045003}
  {\bibfield  {journal} {\bibinfo  {journal} {Rev. Mod. Phys.}\ }\textbf
  {\bibinfo {volume} {93}},\ \bibinfo {pages} {045003} (\bibinfo {year}
  {2021})}\BibitemShut {NoStop}%
\bibitem [{\citenamefont {Changlani}\ \emph {et~al.}(2009)\citenamefont
  {Changlani}, \citenamefont {Kinder}, \citenamefont {Umrigar},\ and\
  \citenamefont {Chan}}]{Changlani2009}%
  \BibitemOpen
  \bibfield  {author} {\bibinfo {author} {\bibfnamefont {H.~J.}\ \bibnamefont
  {Changlani}}, \bibinfo {author} {\bibfnamefont {J.~M.}\ \bibnamefont
  {Kinder}}, \bibinfo {author} {\bibfnamefont {C.~J.}\ \bibnamefont
  {Umrigar}},\ and\ \bibinfo {author} {\bibfnamefont {G.~K.-L.}\ \bibnamefont
  {Chan}},\ }\href {https://doi.org/10.1103/PhysRevB.80.245116} {\bibfield
  {journal} {\bibinfo  {journal} {Phys. Rev. B}\ }\textbf {\bibinfo {volume}
  {80}},\ \bibinfo {pages} {245116} (\bibinfo {year} {2009})}\BibitemShut
  {NoStop}%
\end{thebibliography}%
\end{document}